\documentclass [12pt]{article}
\usepackage{lscape}                                           %
\usepackage[centertags]{amsmath}
\usepackage{amsfonts} \usepackage{amssymb}\usepackage{eufrak}
\usepackage{colordvi}
\usepackage{epsfig}
\usepackage{a4}
\usepackage{cite}
\usepackage{lscape}
\newcommand{\COMMENTO}[1]{}

\newcommand{\A}{\hat{\cal A}}
\newcommand{\B}{\hat{\cal B}}
\newcommand{\C}{\hat{\cal C}}
\newcommand{\D}{\hat{\cal D}}
\newcommand{\calB}{{\cal B}}

\newcommand{\calE}{{\cal E}}
\newcommand{\calG}{{\cal G}}

\newcommand{\Z}{Z}

\newcommand{\alphant}{\alpha}
\newcommand{\alphat}{{ \tilde \alpha}}
\newcommand{\uno}{ \mathbb{I} }

\begin{document}
\begin{titlepage}
\rightline{DSF-32-2007}
\rightline{NORDITA-2007-28} \vskip 3.0cm
\centerline{\LARGE \bf  Wrapped Magnetized Branes:
}\vskip .5cm
\centerline{\LARGE \bf Two Alternative Descriptions?}
\vskip 1.0cm \centerline{\bf P. Di Vecchia
$^a$ $^b$, A. Liccardo $^c$, R. Marotta $^c$,
I. Pesando $^d $ and F. Pezzella $^c$}
\vskip .6cm \centerline{\sl $^a$
The Niels Bohr Institute, Blegdamsvej 17, DK-2100 Copenhagen \O, Denmark}
\vskip .4cm
\centerline{\sl $^b$
Nordita, Roslagstullsbacken 23, SE-10691 Stockholm, Sweden}
\vskip .4cm \centerline{\sl
 $^c$ Dipartimento di
Scienze Fisiche, Universit\`a di Napoli and INFN, Sezione di
Napoli} \centerline{\sl Complesso Universitario Monte
S. Angelo ed. 6, via Cintia,  I-80126 Napoli, Italy}
\vskip 0.4cm \centerline{ \sl $^d$ Dipartimento di Fisica Teorica,
  Universit\`a di Torino and INFN, Sezione di Torino,} \centerline{\sl
 via P. Giuria 1,  I-10125, Torino, Italy}
 \vskip 1cm

\begin{abstract}
We discuss two inequivalent ways for describing
magnetized $D$-branes wrapped $N$ times on a
torus $T^2$. The first one
is based on a non-abelian gauge bundle  $U(N)$, while the
second one is obtained by means of a Narain  T-duality
transformation acting on a theory with non-magnetized branes.
We construct in both descriptions the boundary state
and the open string vertices and show that they give rise
to different string amplitudes. In particular, the description
based on the gauge bundle has open string vertex operators
with momentum dependent Chan-Paton factors.
\end{abstract}
\vfill  {\small{ Work partially supported by the European
Community's Human Potential Programme under contract
MRTN-CT-2004-005104 ``Constituents, Fundamental Forces and Symmetries
of the Universe'' and by the Italian M.I.U.R. under contract
PRIN-2005023102 ``Strings, $D$-branes and Gauge Theories''. Three of us (PDV, RM, FP) thank
the Galileo Galilei Institute for Theoretical Physics for their kind  hospitality and for partial support during the
completion of this work.}}
\end{titlepage}

\newpage

\tableofcontents       %
\vskip 1cm             %

\section{Introduction}
\label{intro}

String theories are described  by two-dimensional conformal field theories
and  are consistent  only in  twenty-six (bosonic string)  or ten
(superstring) space-time dimensions. As it is well-known,
phenomenology imposes all of them to be compactified, except
the four
dimensions observed in our universe. The compactification
procedure, however, always introduces a
certain number of new fields called {\em moduli}
whose expectation values are related, for example,
to the size and the shape of the compact manifold
and determine the parameters of the four dimensional effective Lagrangian.
Without fixing these expectation values, string models are not predictive.
A lot of work has been done in the last few years to fix the value of these
moduli, but we are left with the problem that there are too many
possibilities and there is the feeling
among many string theorists that something
important is still missing.

In general, in order to obtain an ${\cal{N}}=1$ supersymmetric version
of the Standard Model, one needs to compactify the six extra dimensions
in a Calabi-Yau six-dimensional space. However, it is not possible,
in  general, to
have an explicit formulation of string theory in these backgrounds.
Therefore it is not really possible  to construct explicit extensions
of the Standard Model
and compare their results with phenomenology.
If one wants to do that, then it is necessary to
restrict oneself to orbifolds and orientifolds of toroidal compactifications.

Starting from the observation that one wants chiral fermions, as required
by the Standard Model, string theory models based on intersecting branes
have been proposed and extensively studied. In particular, type IIA
orientifolds with intersecting D6 branes, together with their counterparts
in type IIB theory, have provided a phenomenologically
interesting class of very explicit string
compactifications~\cite{0301032,0401156,0502005,0610327}.

In these models
one only considers the simplest case in which
the six-dimensional compact manifold is  the product of three
two-dimensional tori $ T^2  \times T^2  \times T^2 $. If, for the sake
of simplicity,  the analysis is limited
to the torus described
by the two coordinates $x^1$ and $x^2$ with
one stack of D6 branes being placed
along the $x^1$ axis and a different one at an angle
$ \theta$ ($0 \leq \theta\leq\pi $) in the plane $(x^1, x^2)$,
then one gets the
following boundary conditions for an open string having one endpoint attached
to the first stack of branes and the other endpoint
attached to the other stack:
\begin{eqnarray}
\partial_{\sigma} \left[ \cos ( \theta) X^1 -
\sin ( \theta) X^2 \right]_{\sigma=0} =0~~;~~
\partial_{\tau} \left[ \sin ( \theta) X^1 +
\cos ( \theta) X^2 \right]_{\sigma=0} =0
\label{intbou}
\end{eqnarray}
and Neumann and Dirichlet boundary conditions respectively in the
$1,2$ directions at $\sigma = \pi$.

If one considers a squared torus with radii $R_1$ and $R_2$, then
the angle $\theta$ is easily seen to be given by:
\begin{eqnarray*}
\tan \theta = \frac{m R_2}{n R_1}
\label{tantheta}
\end{eqnarray*}
where $(n,m)$ are the wrapping numbers along the directions
$1$ and $2$ respectively of the second stack of branes.
If one performs a T-duality transformation along the direction
$x^2$, that amounts to
exchange $\sigma \leftrightarrow \tau$, the boundary conditions
in Eq. (\ref{intbou}) are transformed into the following ones:
\begin{eqnarray*}
\left[ \partial_{\sigma}  X^1 -
\tan \pi \nu  \partial_{\tau} X^2 \right]_{\sigma=0} =0~~;~~
\left[ \partial_{\sigma }   X^2 +
\tan \pi \nu  \partial_{\tau} X^1 \right]_{\sigma=0} =0
\label{intboub}
\end{eqnarray*}
that are the boundary conditions for an open string having the endpoint
at $\sigma =0$ attached to a brane with a constant
magnetic field, given   by:
\begin{eqnarray*}
\tan \pi \nu  \equiv  2 \pi \alpha' f_{12}
=\frac{m}{n} \frac{\alpha'}{R_1 R_2} ,
\label{F}
\end{eqnarray*}
where $\tan \pi \nu $ is obtained from $\tan \theta$ by the
T-duality transformation: $R_2  \rightarrow \frac{\alpha'}{R_2}$.
In the T-dual theory the integer $n$ multiplies the
volume of the T-dual torus,
and thus it plays the role of the wrapping number
of the brane on the whole torus\footnote{
This wrapping number, which we use in the paper,
 is not the same thing of the geometrical
embedding since it contains less information.
A geometrical embedding of a $T^2$ into a $T^2$
is characterized by a
matrix $\left(\begin{array}{cc} p & j \\ 0 & q\end{array}\right)$, up to
  $SL(2,Z)$ transformations, which has wrapping $n=p q$.

}.

Being the brane compactified on a torus $T^2$, the
first Chern class for an $SU(N)$
non-abelian gauge field living on it must be an integer, i.e.:
\begin{eqnarray}
\int_{{\cal M}} Tr \left( \frac{F}{2 \pi} \right)
= 2 \pi \alpha' F_{12} N  = m'
\Longrightarrow   2 \pi \alpha' F_{12} = \frac{m'}{N}
\label{chernc89}
\end{eqnarray}
being ${\cal M}$ the brane worldvolume.
It is easily seen that the wrapping number
$m$, along the direction $x^2$
in which the T-duality transformation is performed, becomes the
magnetic charge $m'$ and the wrapping number $n$ along the
other direction $x^1$ becomes the rank of the
gauge group $N$ that is also equal to the wrapping number on
the entire torus $T^2$. In conclusion, this analysis suggests that a brane
wrapped $n$ times along the first cycle of the torus and $m$ times
along the second cycle of the torus becomes, under a T-duality performed
along the second axis $x^2$,  a magnetized brane with magnetic flux $m$
that is wrapped $n$ times along the whole torus and that is described by a
non-abelian gauge theory with gauge group $U(N)$ with $N=n$.

The fact that, in order to describe branes wrapped $N$ times
on the torus $T^2$, one needs a non-abelian gauge theory $U(N)$,
has been advocated by many
authors~\cite{Taylor:1997dy,
Hashimoto:1997gm,Guralnik:1997sy,Hashimoto:1996pd,Rabadan:2001mt}.
In particular, in Ref.~\cite{Cremades:2004wa}, it has
been used  for computing, among other things, the Yukawa
couplings in the
field theory limit ($\alpha' \rightarrow 0$) corresponding to massless
open string states attached to magnetized branes. In this approach
a $D$-brane wrapped $N$ times along the torus $T^2$ is described
by a $U(N)$ gauge theory just like a stack of $N$ $D$-branes only wrapped
once.
The difference between the two systems is that
in the latter case the gauge holonomy is the identity, while
it is not trivial in the former.
This implies that for $N$ branes wrapped once
the gauge theory quantities are periodic
in going around the two cycles of the torus, while
for an $N$-tuply wound D brane one gets
instead a gauge bundle whose non gauge invariant
quantities are periodic only up to a
gauge transformation.

At this point a question is natural: do the previous considerations mean that
a brane wrapped $N$ times around  the torus $T^2$ is necessarily
described by a gauge bundle of a $U(N)$ gauge
theory? Instead of trying to answer this question
directly, let us observe that,
if one uses an abelian rather than a non-abelian field, then
the factor  $N$ in Eq. (\ref{chernc89})
could be in principle reproduced in a different way. In fact, in this case,
Eq. (\ref{chernc89}) becomes:
\begin{eqnarray}
\int_{\cal{M}}  \left( \frac{F}{2 \pi} \right)  = 2 \pi \alpha' F_{12} N  = m'
\label{chernc89a}
\end{eqnarray}
where now  the factor $N=n$ is not  obtained from the trace
over the non-abelian group as before, but
from the fact that the brane worldvolume is
$N$ times the volume of the torus
$(2 \pi \sqrt{\alpha'} )^2$, differently from
the non-abelian case where the brane worldvolume coincides with the torus, as a consequence
of the different periodicity conditions on the embedding coordinates.

This second logical possibility can be realized, for instance,
by a particular system of $D$-branes that
we call {\em Narain branes}. These are indeed obtained
from a system of non-magnetized branes by acting with the Narain T-duality group.
It seems to us that this point of view has been taken in
Ref.s~\cite{9810072,Blumenhagen:2000fp,Blumenhagen:2000wh, Bianchi:2005yz, Antoniadis:2006, Antoniadis:2007}.
This was also the point of view taken in
Ref.s~\cite{Pesando:2005df,DiVecchia:2006gg}, but with some differences
with respect to the previous ones.

In this paper we  will discuss  both points of views and compare
them. There is also another reason for our analysis. In fact, while
on the side of the intersecting branes  a unique and complete string
description (up to non geometrical data) is available and amplitudes involving both twisted and
untwisted open strings have been computed, on the side of magnetized
branes, instead, only  partial tree-level string calculations have
been performed and, as far as we can see, a complete string
description of magnetized branes is still lacking. For example, in
the case of the Yukawa couplings, only a part of the amplitude has
been computed at tree-level in string
theory~\cite{0610327,BERTOLINI}, while the rest has been obtained in
the field theory limit~\cite{Cremades:2004wa}. The complete
expression of the Yukawa couplings has been obtained from the
corresponding computations in the intersecting branes scenario via
T-duality \cite{0302105,0303083,0303124} or from the two-loop
twisted partition function \cite{russciu,duo}.

In this paper we make the first step toward a more complete
string theory formulation of magnetized branes on the torus.

In particular, on the one hand, we show that, in order to
describe wrapped branes,  we need to extend the
concept of gauge bundle to string theory and in this framework
we write the equations
that characterize the physical states when one goes  around
the two one-cycles of the torus.
We  show that in this case the Chan-Paton factors,
unlike the usual ones describing the
non-abelian degrees of freedom, are momentum dependent. Then we construct
the boundary state corresponding to
wrapped magnetized branes described by gauge bundles and
compute  the one-loop partition function. The boundary state is
constructed in two different ways.
The first consists in factorizing the annulus diagram computed in the
open string channel which
fixes the boundary state up to a
phase factor and the second in a direct calculation
involving the non-abelian Wilson loop.
We find agreement between the two procedures up to a phase factor.

On the other hand,
following the other logical possibility discussed above for describing
the wrapped branes, we start from a theory with no  gauge field and
we get a theory with a non-vanishing gauge field  by using  the
general T-duality group found by Narain.
We then  apply the same technique based on T-duality for
constructing a boundary state
with a gauge field on it from the usual boundary state. It turns out that the
 boundary states determined in the two approaches
 are not identical, but give the same one-loop amplitude.
The boundary state obtained in the gauge bundle scheme contains an
extra phase factor that, however, does not contribute to the annulus
diagram.

We then compute in these two theories disk amplitudes
involving both open and closed strings showing that they
are different. Hence, if we only  focus on the first
Chern class constraint in order to characterize the T-dual of
intersecting branes, we do not have yet
enough elements for distinguishing which of these two inequivalent
descriptions is the right one.

The paper is organized as follows. In Sect. \ref{sect-review} we study open
strings in an arbitrary toroidal background interacting with a magnetic field
living on the compactified directions. We introduce the conserved generalized translation
operator and the notion of gauge bundle.

In Sect. \ref{sect-nonabelian-open} we discuss the gauge bundle
in string theory as a description of wrapped magnetized
space-filling  branes.   We construct
the corresponding boundary state, compute the one-loop diagram, and
give the open string vertices containing the Chan-Paton factors that are
momentum dependent.
In this paper we will call {\em non-abelian branes} those based on a non-abelian gauge bundle.

In Sect. \ref{sect-narain-closed} we discuss
the Narain branes.  In particular, we construct
their boundary state, that turns out to be equal to that of the
non-abelian branes apart from a phase factor, and their open
string vertices. We show that  the two kinds of
branes, even if they  have the same free-energy, are
indeed different objects because they
have a different boundary state and different scattering amplitudes
involving both open and closed strings.

Many of the calculations are presented in four Appendices.
In Appendix  \ref{app-conventions} we summarize our conventions.
Appendix \ref{sect-reviewA} is devoted to the
solution of the equations of motion
of open and closed strings in closed and open string backgrounds.
In Appendix \ref{T-duality} we discuss the
transformations of various quantities
under the general Narain group of T-duality and in Appendix D we perform the
path-ordering calculation of the boundary state with a background gauge field.

\section{Open strings in flux backgrounds}
\label{sect-review}

In this section we study the effects of turning on a background gauge field
living on a ${\hat d}$-dimensional torus
and interacting with closed string backgrounds. First we review the solution of the equations of motion for open strings (the case of closed strings is discussed in Appendix \ref{sect-reviewA}) and
then we analyse how the translation generator gets modified when a  magnetic field is turned on.

\subsection{Open string in open and closed string background.}
\label{open1}

Let us consider open
strings on the ${\hat d}$-dimensional torus, interacting with
constant gravitational and  Kalb-Ramond  backgrounds
and with an open string background consisting of two abelian gauge fields with
constant field strengths $F^{(0)}_{ij}$ and $F^{(\pi)}_{ij}$  acting
at the two end-points  $\sigma =0, \pi$ of the string and, in
general, independent of each other. Such a system is described by the following action~\footnote{
With respect to the notation used in \cite{DiVecchia:2006gg} we have
set $q_\pi\rightarrow -q_\pi$.}:
\begin{eqnarray}
S = S_{bulk} + S_{boundary}
\label{spluss}
\end{eqnarray}
where $S_{bulk}$ is given by:
\begin{eqnarray}
S_{bulk} = - \frac{1}{4 \pi
  \alpha'}
\int d\tau \int_{0}^{\pi} d \sigma
\left[ G_{ab} \partial_{\alpha} X^a \partial_{\beta} X^b \eta^{\alpha \beta}
     - B_{ab} \epsilon^{\alpha \beta} \partial_{\alpha} X^a
\partial_{\beta} X^b \right] \label{acti853}
\end{eqnarray}
being the world-sheet metric $\eta_{\alpha \beta} =
\mbox{diag}(-1,1)$ and
$\epsilon^{01} =1$, while   $S_{boundary}$ is:
\begin{eqnarray}
S_{boundary}  \!\!\!&=& - q_0 \int d \tau A_{i}^{(0)} \partial_{\tau} X^{i}
|_{\sigma =0} + q_{\pi} \int d \tau A_{i}^{(\pi)} \partial_{\tau} X^{i}
|_{\sigma =\pi} \nonumber \\
\!\!\!&=& \frac{q_0}{2} \int d \tau F_{i j}^{(0)} X^j
  {\dot{X}}^{i}|_{\sigma =0}
-\frac{q_{\pi}}{2} \int d \tau F_{i j}^{(\pi)} X^j
  {\dot{X}}^{i}|_{\sigma=\pi}
~~;~~
i,j=1,...,\hat{d}
\label{s2}
\end{eqnarray}
where $q_0$ and $q_{\pi}$ are the charges located at the two
end-points of the open string. In Eq. (\ref{s2}) we have used
the following form for the background gauge fields
\begin{eqnarray}
A_{i} =-\frac{1}{2} F_{i j} x^j~.
\label{A-F-main}
\end{eqnarray}
The field $A_{i}$ in Eq. (\ref{A-F-main})
is not a periodic quantity
when we go around one of the two one-cycles of the torus.
However, on the torus, gauge non-invariant
quantities as $A_i$ have only to be periodic up to a gauge
transformation \cite{Hooft81}:
\begin{eqnarray}
A_i(x^j + 2\pi \sqrt{\alpha'} \delta^j_{l} )
=
\Omega_{l}(x)~A_i(x^j) ~\Omega_{l}^{-1}(x)
-i \frac{1}{q} \Omega_{l}(x) ~\partial_i \Omega_{l}^{-1}(x)
\label{Omega-main}
\end{eqnarray}where $q$ is the gauge coupling constant and $\Omega_l(x) \equiv\Omega_l(x^{j \neq l})$ is the gauge transition function.
{From} now we mean by {\em gauge bundle} the
assignment of a background field, together
with a transition function  which fixes the periodicity property of the gauge field.
Analogously, matter fields in the adjoint representation have to satisfy the periodicity conditions
\begin{eqnarray}
\Phi(x^j + 2\pi \sqrt{\alpha'} \delta^j_{l} )
=
\Omega_{l}(x)~\Phi(x^j) ~\Omega_{l}^{-1}(x).
\label{Omega-sect}
\end{eqnarray}
Notice that, if we perform  a gauge transformation
\begin{eqnarray*}
A_i^\omega(x)
&=&
\omega(x)~A_i(x) ~\omega^{-1}(x)
-\frac{i}{q}\omega(x) ~\partial_i \omega^{-1}(x)
\label{gauge}
\end{eqnarray*}
the transition functions transform  as
\begin{eqnarray}
\Omega^\omega_j(x)=
\omega(x^1, \dots, x^j+2\pi\sqrt{\alpha'},\dots, x^{\hat d}) ~\Omega_j(x) ~\omega^{-1}(x^1,\dots, x^{\hat d})\label{ogt}.
\end{eqnarray}
Furthermore, they also have to satisfy the {\em cocycle condition}, which
simply means that the gauge fields must be unchanged when
translated along a closed path:
\begin{eqnarray}
\Omega_{j}(x^k+2\pi\sqrt{\alpha'}\delta^k_{i}) \Omega_{i}(x^{k})\Omega_{j}^{-1}(x^{k})\Omega_{i}^{-1}( x^k+2\pi\sqrt{\alpha'}\delta^k_{j})=\mathbb{I}_{ N}\label{cocycle}.
\end{eqnarray}
For the choice of the gauge field given in Eq. (\ref{A-F-main}), the gauge transition functions satisfying the identity in Eq. (\ref{Omega-main}) are
\begin{eqnarray}
\Omega_i(x) = e^{-i \pi \sqrt{\alpha'} q F_{ i j} x^j}.
\label{Omega1}
\end{eqnarray}
With the previous choice the cocycle condition is trivially satisfied because it is equivalent to require that the first
Chern class is an integer, as we will show shortly.

The above considerations can be extended to the case of a $U(1)$
gauge field that is included in a $U(N)$ gauge theory. In this case we
have a non-abelian gauge bundle
where the gauge transition functions are non-trivial unitary matrices.
A possible choice for them is \cite{Hooft81}:
\begin{eqnarray*}
\Omega_j=e^{-i \pi \sqrt{\alpha'} q F_{ j i} x^i}\omega_j
\end{eqnarray*}
where we have extracted the $U(1)$ factor from the gauge transition functions.
The cocycle condition imposes the following constraints on the  $\omega$'s:
\begin{eqnarray}
\omega_i ~\omega_j=
e^{-i 2\pi q \hat F_{ ij} } ~\omega_j ~\omega_i~~;\qquad
{\hat{F}}_{ ij} \equiv 2 \pi \alpha' F_{ij}.
\label{omega-Td1}
\end{eqnarray}
By taking the determinant of the previous expression
it follows that the field
strength must satisfy the condition:
\begin{eqnarray}
q {\hat{F}}_{ij}~ N=n_{ij} \in \mathbb{Z}\label{fcc}
\end{eqnarray}
where $F_{ij}$ is constant and $n_{ij}$ is an integer.
Eq. (\ref{fcc}) is indeed  satisfied because it coincides with the
first Chern class for a non-abelian gauge field, that must be an integer.
In the case of an $SU(N)$ gauge group, one can explicitly
construct the matrices $\omega_i$  in terms of the constant 't Hooft
matrices $P_N$ and $Q_N$ \cite{Hooft81} given in Appendix A:
\begin{eqnarray}
\omega_i\equiv P_{N}^{ s_{i}}
Q_{ N}^{ t_i}\qquad i=1,\dots, {\hat{d}} \label{thooft}
\end{eqnarray}
with $s_i$ and $t_i$ suitable integers. Since $P_N Q_N=e^{2\pi i /N} Q_N P_N$, then
the cocycle condition written in Eq. (\ref{omega-Td1}) can be easily satisfied by choosing:
\begin{eqnarray*}
 n_{ij} =
s_{i} t_{j} - s_{j} t_{i} .
\label{co13}
\end{eqnarray*}
with $n_{ij}$ defined in Eq. (\ref{fcc}).
One could also add Wilson lines and this can be done in two different
ways:  either  by adding  them as a constant in the expression of the
gauge field or in the transition functions. We will
discuss these two possibilities later on.

{From} the action in Eq. (\ref{spluss}) one can derive the
equations of motion in the bulk given by:
\begin{eqnarray}
\partial_{\alpha} [ G_{ij} \partial^{\alpha} X^j ] =0
\label{bulkequa}
\end{eqnarray}
and the two boundary conditions at $\sigma =0, \pi$:
\begin{eqnarray}
\left[ G_{ij} \partial_{\sigma} X^j + ( B_{ij} - 2 \pi \alpha' q_{0,\pi} F_{ij}^{(0,\pi)})
  \partial_{\tau} X^j \right]_{\sigma =0,\pi} =0.
\label{bou0}
\end{eqnarray}
The solution of these equations can be easily found in the simplest case in which
\begin{equation*}
q_{0}=q_{\pi}=q \,\,\,\,\,\,\,\,\,\, F_{ij}^{(0)}=F_{ij}^{(\pi)}=F_{ij}.
\end{equation*}
and the following condition holds:
\begin{equation*}
\mbox{det} (q_{0}F^{(0)} - q_{\pi}F^{(\pi)})_{ij}=0
\end{equation*}
corresponding to the so-called {\em dipole} string.
The case in which this determinant is different from zero
corresponds
to the {\em dycharged}  string.

In the dipole case the  general solution~\cite{CHU,Chu:2000wp} can
be written as (see Appendix \ref{dettagli} for details):
\begin{eqnarray}
X^i(\sigma,\tau) &=&
\frac{1}{2}\left[ \hat X_L^i(\tau+\sigma)+ \hat X_R^i(\tau-\sigma) \right]
\label{icso}
\end{eqnarray}
where the left and right moving parts are given, up to a constant, by:
\begin{eqnarray}
\hat X_L^i(\tau+\sigma)
&=&
(G^{-1} {\cal E})^i_j
\left(
X_{L }^j(\tau+\sigma)
\right)
\label{icsoaa}
\end{eqnarray}
and
\begin{eqnarray}
\hat X_R^i(\tau-\sigma)
&=&
(G^{-1} {\cal E}^T)^i_j
\left( X_{R }^j(\tau-\sigma)
\right)
\label{icsob}
\end{eqnarray}
with $\calE$ defined by:
\begin{eqnarray}
{\cal E} &=& \parallel \calE_{i j} \parallel=E^T + 2\pi \alpha' q_0 F
\equiv G -{\cal B}
\label{calE}
\end{eqnarray}
being
\begin{eqnarray}
E &=& \parallel E_{i j} \parallel = G+B
\nonumber
\end{eqnarray}
and
\begin{eqnarray*}
 X_{L }^i(\tau+\sigma)
&=&
x^i +2\alpha'  {\cal G}^{ij}p_{j} (\tau+\sigma)
+ i \sqrt{2\alpha'} \sum_{n\ne 0} \frac{\alpha_n^i }{n}
e^{-i n(\tau+\sigma)}
\nonumber\\
 X_{R}^i(\tau-\sigma)
&=&
x^i +2\alpha'  {\cal G}^{ij}p_{j} (\tau-\sigma)
+ i \sqrt{2\alpha'} \sum_{n\ne 0} \frac{{\alpha}_n^i }{n}
e^{-i n(\tau-\sigma)}
\end{eqnarray*}
where ${\cal{G}}_{ij} $ is the {\em open string metric}:
\begin{eqnarray*}
\calG
\equiv \calE^T G^{-1} \calE
\label{osm}
\end{eqnarray*}
and  ${\cal{G}}^{ij}$  is its inverse.
The quantization of the theory requires the following commutation
relations~\cite{CHU,Chu:2000wp,Chu:2005ev}:
\begin{eqnarray*}
[x^i, x^j]= i ~2\pi\alpha' \Theta^{i j}
~~~~
[x^i, p^j]= i {\calG}^{i j}
~~~~
[\alpha^i_m,\alpha^j_n] = {\calG}^{i j} m \delta_{n+m,0} \label{comm}
\end{eqnarray*}
where $\Theta$ is defined by the relation:
\begin{equation}
({\cal E}^{-1})^{ i j} = {\cal G}^{i j} -\Theta^{i j}.
\end{equation}
The operator $L_{0}$ is given by the Hamiltonian in Eq. (\ref{hamilto}) and a straightforward calculation gives:
\begin{eqnarray}
L_0 = \alpha'  p_i {\cal{G}}^{ij} p_j +  \sum_{n=1}^{\infty}
{\cal{G}}_{ij} \alpha^{i}_{-n} \alpha^{j}_{n}.
\label{LO}
\end{eqnarray}

\subsection{Translation generator in the presence of a magnetic field.}
\label{sect-spectrum-p}

In our compactified string theory, described by
the bulk and boundary actions respectively given in Eq.s  (\ref{acti853}) and (\ref{s2}), the conjugate momentum density given by:
\begin{eqnarray*}
P_i
&=&
\frac{1}{2 \pi \alpha'  }
\left[ G_{ij} {\dot{X}}^j + B_{ij} X^{'j} \right]  \nonumber\\
&& +\frac{1}{2 }q_0 F_{i j}^{(0)} X^{j}(0) ~\delta(\sigma)
-\frac{1}{2 }q_\pi F_{i j}^{(\pi)} X^{j}(\pi) ~\delta(\pi-\sigma)
\label{gau-var-P}
\end{eqnarray*}
is not gauge invariant, because of the gauge choice
made in Eq. (\ref{A-F-main})
and is not a  conserved charge as in the case with $F=0$.
This is not  surprising   because the string action, in
the presence of a magnetic field,  is not invariant under translations.
However, one can easily see  that the string action is
indeed invariant under a suitable combination of a  translation and a
gauge transformation~\cite{0401040} .
In particular, with the gauge chosen in Eq. (\ref{A-F-main}), the action
is invariant under the combination of a translation
$X^i\rightarrow X^i+\epsilon^i$  (under which the gauge field transforms as
$A_i \rightarrow A_i+\partial_jA_i\epsilon^j$) and
a gauge transformation $A_i\rightarrow A_i-\partial_i\phi$
with $\phi=\frac{1}{2}F_{ij}X^j\epsilon^i$. We will refer to this transformation as a {\em generalized translation}.
The N\"oether current associated to such an invariance
is given by:
\begin{eqnarray*}
J^{\alpha}_i(\tau,\,\sigma) =-
\frac{1}{2\pi\alpha'}[G_{ij}\partial^{\alpha} X^j -B_{ij}\epsilon^{\alpha\beta} \partial_\beta X^j]
\end{eqnarray*}
which is conserved as a consequence of the equations of motion.

The conservation of the previous current implies  \cite{PRD5249}:
\begin{eqnarray*}
0&=&\int_0^\pi d\sigma \partial_\alpha J_{i}^{\alpha} =\int_0^\pi d\sigma\, \partial_\tau J^0_i(\sigma)+ \left[ J^1_i(\sigma)|_{\sigma=\pi}- J^1_i(\sigma)|_{\sigma=0} \right]  =\nonumber\\
&=&\partial_\tau \left[\int_0^\pi d\sigma \frac{1}{2\pi\alpha'}[G_{ij}\dot{X}^j+B_{ij}{X^j}']
+q_0F_{ij}^{(0)} X^j|_{\sigma=0}- q_\pi F_{ij}^{(\pi)} X^j|_{\sigma=\pi}\right]
\end{eqnarray*}
where we have used the open string boundary conditions
given in Eq.s (\ref{bou0}).
It follows that the  generator of such a transformation is
a constant of the motion  that  is simply given by
\cite{0401040,PRD5249}:
\begin{eqnarray}
\hat T_i&=&
\int_0^\pi \!d\sigma\,
\left\{ \frac{1}{2\pi \alpha'}[G_{ij}\,\dot{X}^j \,+ B_{ij}\, {X^j}']+q_0F^{(0)}_{ij}\,X^j\delta(\sigma)-q_\pi
F^{(\pi)}_{ij}\,X^j\delta(\sigma-\pi)  \right\}\nonumber\\
&=&(q_0F^{(0)}-q_\pi F^{(\pi)})_{ij}x^j+\delta_{q_0F^{(0)}-q_\pi
F^{(\pi)};0}\,{\cal G}_{ij}p^j.     \label{gtg}
\end{eqnarray}
It differs  from the conjugate momentum for the one half factor in front of the terms depending on the gauge field. Moreover it  satisfies the following commutation relation \cite{PRD5249}:
\begin{eqnarray}
\left[ \hat T_i, \hat T_j\right]= i\, \left( q_0F^{(0)}-q_\pi
F^{(\pi)} \right)_{ij}.\label{cr}
\end{eqnarray}
It is interesting to observe that, when the dipole condition is satisfied,  the same expression for the translation operator could be obtained by considering a slight modification of Eq. (\ref{s2}):
\begin{eqnarray}
\hat S_{boundary}
&=&  -q\int  d \tau  \int_{0}^{\pi} d\sigma ~F_{i j} X^{' j} {\dot{X}}^{i}
\nonumber\\
&=& - q\int  d \tau \int_{0}^{\pi} d\sigma  ~\left[\partial_\sigma
\left(\frac{1}{2 }F_{i j} X^{j} \dot{X}^i \right)
-\partial_\tau\left( \frac{1}{2 }qF_{i j} X^{j} X^{' i} \right)\right].
\label{s2hat}
\end{eqnarray}
This expression is equivalent to Eq. (\ref{s2}) up to a total derivative with respect
to $\tau$ and is gauge invariant.
The conjugate momentum computed from the bulk action in Eq. (\ref{acti853})
with the addition of the boundary term in Eq. (\ref{s2hat}) turns out to be :
\begin{eqnarray}
\hat P_i
&=&
\frac{1}{2 \pi \alpha'  }
\left[ G_{ij} {\dot{X}}^j + B_{ij} X^{'j} \right]
-q F_{i j} X^{' j}
=
\frac{1}{2 \pi \alpha'  }
\left[ G_{ij} {\dot{X}}^j + ( B_{ij} - q\hat F_{i j} )X^{'j} \right]
\nonumber\\
&=&
\frac{1}{2 \pi \alpha'  }
\left[ (\calE^T)_{ij} \partial_+ X^j + (\calE)_{ij} \partial_- X^j \right].
\label{gau-inv-P}
\end{eqnarray}
When integrated in  $d\sigma$, it yields precisely  Eq. (\ref{gtg})
for the dipole case.

On a torus the system must be invariant under a generalized
discrete translation, i.e. $x^i \rightarrow x^i + 2 \pi \sqrt{\alpha'}$. This means that the physical states
are those which satisfy the following identity:
\begin{equation}
{\cal T}_i  |phys\rangle \equiv e^{i 2\pi \sqrt{\alpha'}  \hat T_i} |phys\rangle
= |phys\rangle.
\label{u1-phys-period-inv}
\end{equation}
In particular, when the dipole condition is satisfied,  the translation operator
becomes identical to  the momentum operator in Eq. (\ref{gau-inv-P}) (as it follows from
Eq. (\ref{gtg}))
whose spectrum is determined by imposing the previous constraint.
In this way  one gets:
\begin{equation*}
 p_i |k\rangle  =  k_i |k \rangle
= \frac{n_i}{\sqrt{\alpha'}}
|k \rangle
\end{equation*}
where $k_i$ is the eigenvalue of the operator $p_i$.
Taking into account these eigenvalues one can rewrite the operator
$L_{0}$ in Eq. (\ref{LO}) as follows:
\begin{equation}
L_{0} = n_i
{\cal{G}}^{ij} n_j + \sum_{n=1}^{\infty}
{\cal{G}}_{ij} \alpha^{ i}_{-n} \alpha^{j}_{n} .
\label{L0ni62}
\end{equation}
The previous analysis is valid when we have just $D$-branes wrapped
once on the torus  and having a $U(1)$
background gauge field turned-on on their worldvolume.
In the next section we will generalize the previous construction
to the case of $D$-branes wrapped $N$ times.

\section{Non-abelian branes}
\label{sect-nonabelian-open}
In this section we provide a description of $D$-branes wrapped
$N$ times along the two-cycles of the torus $T^2$. We start
discussing in  sect. \ref{F=0}
the case with vanishing gauge field that has been already discussed
in the literature and then we analyze in sect. \ref{Fneq0}
the case with $F \neq 0$.
We show that, in this description,
$D$-branes wrapped $N$ times along the torus $T^2$
have Chan-Paton factors that are momentum dependent.  Then in
sect. \ref{nondege} we discuss the non-degenerate case and in
sect. \ref{dege} the degenerate one. Sect. \ref{boundary} is devoted
to the construction of the boundary state and to the one-loop
annulus diagram. Finally the open tachyon vertex is
constructed in sect. \ref{strive}.

\subsection{Non-abelian gauge bundle: $F=0$}
\label{F=0}

Let us start discussing  the case with $F=0$,  but with $A$ having a non trivial background value or, equivalently, when a Wilson line background is turned on.
This is the case considered in Ref.~\cite{Hashimoto:1996pd} where the perturbative dynamics of open strings attached to multiply wound $D$-branes is analyzed.
A very important point made in  Ref.~\cite{Hashimoto:1996pd}  is that a $D$-brane wrapped $N$ times around a torus can be
described as a brane with gauge group $U(N)$ and with
a non trivial holonomy group  ${\cal H}\neq\uno$.
Actually, as stressed in Ref.~\cite{Polchinski9602052}, the
existence of a non-trivial holonomy is what makes the difference
between a bound state of $N$ $D$-branes each wrapped
once around the torus (such a state has gauge group $U(N)$ but
trivial holonomy  ${\cal H}=\uno$) and a single N-tuply wound $D$-brane.
On the other hand, a non-trivial gauge holonomy ${\cal H}$ arises when a Wilson line background is turned on, being
${\cal H}=P[e^{-iq\oint A}]$ and $P$ stands for the path ordering.
Thus Wilson lines provide a natural way to describe multiply wound $D$-branes.
However, as discussed in Ref.~\cite{Hashimoto:1996pd}, a Wilson line background implies non standard kinetic terms for the fields. Therefore it is useful to make a field redefinition which actually exchanges
the Wilson line background with non-trivial boundary conditions.
More explicitly, let us consider the following $U(N)$ Wilson lines background:
\begin{equation*}
A_i= \Theta_i= \left(\begin{array}{ccc}
a^{1}_i & \dots & 0 \\
0 & \ddots & \\
0 & \dots & a^{ N}_i
\end{array}
\right).
\label{Wilson1}
\end{equation*}
It implies the existence of a non-trivial holonomy group:
\begin{equation}
{\cal H}=P[e^{-iq\oint A}]=e^{-iq 2\pi \Theta_i{\sqrt{\alpha'}}}.
\label{holon}
\end{equation}
Being the gauge field background $A_i$ constant, the
periodicity condition in Eq. (\ref{Omega-main}) is simply
satisfied by the gauge transition function
$\Omega_l=\uno_N$, which means that
both the gauge and the matter fields have trivial boundary conditions:
\begin{eqnarray}
A_i(x^j + 2\pi \sqrt{\alpha'} \delta^j_{l} )
=
A_i(x^j) \qquad
\Phi(x^j + 2\pi \sqrt{\alpha'} \delta^j_{l} )
=
\Phi(x^j) .
\label{Omega-trivial}
\end{eqnarray}
However, the non-zero value of the background
field $A$ generates non-standard kinetic terms.
It is therefore useful to perform a gauge transformation with the gauge function
$\omega(x^1,\dots, x^{\hat d})=e^{-iq \Theta_i x^i}$
so, according to Eq. (\ref{ogt}),  one gets:
\begin{equation}
A^{\omega}_i= 0 ~~~~ \Omega_i^{\omega}= e^{-iq 2\pi \Theta_i{\sqrt{\alpha'}}} \label{Wilson2}
\end{equation}
corresponding to a new description in which the Wilson
line background is zero, but the gauge and matter fields satisfy the non-trivial
boundary conditions given in
Eq.s (\ref{Omega-main}), (\ref{Omega-sect})
with $\Omega$ given in (\ref{Wilson2}).
Notice that the transition function in Eq. (\ref{Wilson2}) coincides with the holonomy matrix in Eq. (\ref{holon}).

\subsection{Non-abelian gauge bundle in string theory: $F \neq 0$}
\label{Fneq0}

In this subsection we discuss the case $F\neq 0$ and extend the notion of gauge bundle, discussed
in  sect. \ref{open1}, to the string level in the
non-abelian case.
The main difference which occurs in the $F\neq 0$ case is
that, by choosing for example the gauge field as in Eq. (\ref{A-F-main}),
the gauge transition functions are forced to be non trivial, because
$A_i$ itself is not single valued under
$x^i\rightarrow x^i+2\pi\sqrt{\alpha'}$. In other words
with $F\neq 0$ one is forced to have non trivial holonomy.

Let us first go back
to  the abelian case describing a D brane  only wrapped
once on the torus $T^2$. The basic  ingredient
is the identification of the physical states under combined translations and gauge transformations as expressed in Eq. (\ref{u1-phys-period-inv}).
In addition  we have also to impose the consistency condition which requires
the string states to be left invariant when translated along a
closed path. This means that:
\begin{eqnarray}
{\cal T}^{(i)}_{2\pi\sqrt{\alpha'}}
{\cal T}^{(j)}_{2\pi\sqrt{\alpha'}}|phys.\rangle=
{\cal T}^{(j)}_{2\pi\sqrt{\alpha'}}
{\cal T}^{(i)}_{2\pi\sqrt{\alpha'}}|phys.\rangle\label{stco}
\end{eqnarray}
or equivalently:
\begin{eqnarray}
\left[{\cal T}^{(i)}_{2\pi\sqrt{\alpha'}},\,{\cal T}^{(j)}_{2\pi\sqrt{\alpha'}}\right]=0 \label{stcocycle}.
\end{eqnarray}
 Eq. (\ref{stcocycle}) can be considered as  the extension to the string level
 of the cocycle condition given in Eq. (\ref{cocycle}).
 In the abelian case the
 constraint in  Eq. (\ref{stcocycle}) is always  verified. In particular, when the  dipole condition holds, it
 is satisfied because the translation generators commute, as it follows from Eq. (\ref{cr}).
In the dycharged string case, the right hand side of
Eq. (\ref{cr}) is not vanishing anymore and  Eq. (\ref{stcocycle})
imposes the following constraint:
\begin{eqnarray}
 (2\pi\sqrt{\alpha'})^2(q_0F^{(0)}-q_\pi F^{(\pi)})_{ij}=
 2\pi (n^{(0)}_{ij} - n^{(\pi) }_{ij}) \,\,\,\,\,\,  n_{ij}\in\mathbb{Z}
\label{cherncla}
\end{eqnarray}
which is indeed satisfied because this is nothing but the
definition of the first Chern class for a constant field strength.
This is another evidence that $\cal T$ is the right translation
operator.

The previous analysis is valid in the abelian case in which we have
just two branes wrapped once on the torus with two different $U(1)$
background gauge fields turned-on  on their worldvolume. But if we
want to describe two stacks of $D$-branes wrapped respectively $N_{0}$
and $N_{\pi}$ times on the torus, we need to extend the previous
considerations to a non-abelian case where the string states are
dressed with Chan-Paton factors. The gauge group  is $U(N_{0})\times
U(N_{\pi})$. In this case the generator of generalized translations
acts also on the Chan-Paton factors.  We denote by
$\omega_i^{(0,\pi)}$  the matrices acting on them. Physical states
must then satisfy the identification
\begin{eqnarray}
 e^{2\pi \sqrt{\alpha'}\,i \hat T_i} (\omega^{(0)}_i)_{\ell h}
 ({\omega^{(\pi)}_i}^\dag)_{tm} |\Phi,\,h \,t\rangle \equiv |\Phi, \, \ell \,m\rangle,
 \label{bst1}
 \end{eqnarray}
where the background gauge field has been taken in the $U(1)$  part of the gauge group. Moreover
the following cocycle condition has to be satisfied:
\begin{eqnarray}
&&e^{(2\pi\sqrt{\alpha'})i \hat T_j} e^{(2\pi\sqrt{\alpha'})i \hat T_i}
 (\omega^{(0)}_i\,\omega^{(0)}_j)_{\ell h}
 ( {\omega^{(\pi)}_i} \,{\omega^{(\pi)}_j} )_{tm}^\dag
 |\Phi, \,h \,t \rangle\nonumber\\
 &=&e^{(2\pi\sqrt{\alpha'})i \hat T_i}e^{(2\pi\sqrt{\alpha'})i \hat T_j}
 (\omega^{(0)}_j\,\omega^{(0)}_i)_{\ell h}
 ( {\omega^{(\pi)}_j}\,{\omega^{(\pi)}_i})_{tm}^\dag|\Phi,\,h\,t\rangle .
 \label{cocy4}
\end{eqnarray}
This equation is satisfied if we impose the relations:
\begin{eqnarray}
\omega^{(0)}_i\,\omega^{(0)}_j= e^{-2\pi i
\frac{n^{(0)}_{ij}}{N_0}}\omega^{(0)}_j\,\omega^{(0)}_i~~;~~
\omega^{(\pi)}_i\,\omega^{(\pi)}_j= e^{- 2\pi i
\frac{n^{(\pi)}_{ij}}{N_{\pi}}}\omega^{(\pi)}_j\,\omega^{(\pi)}_i.
\label{rc}
\end{eqnarray}
Eq. (\ref{rc}) is the string realization of the constraint given in
Eq. (\ref{omega-Td1}) and again it can be satisfied by taking for
the $\omega$'s the matrices  in Eq. (\ref{thooft}).

Eq. (\ref{bst1}) can also be written in the following suggestive
form:
\begin{eqnarray*}
 e^{2\pi \sqrt{\alpha'}\,i {\hat T}_i}  [ (\omega^{(0)}_i)
 \Lambda ({\omega^{(\pi)}_i}^\dag)]_{\ell m} \lim_{z \rightarrow 0}
  V(z )|0\rangle
 \equiv \Lambda_{\ell m}  \lim_{z \rightarrow 0} V(z)| 0\rangle
 \end{eqnarray*}
where $V(z)$ is the  vertex operator which creates the corresponding  physical state by acting on the conformal
vacuum and the matrix $\Lambda$ is
the Chan-Paton
factor.  The previous  equation implies  the important relation:
\begin{eqnarray}
  e^{2\pi \sqrt{\alpha'}\,i {\hat T}_i}\omega^{(0)}_i
  \Lambda V(z )( e^{2\pi \sqrt{\alpha'}\,i {\hat T}_i} {\omega^{(\pi)}_i})^\dag
    \equiv  \Lambda V(z)
 \label{vertexc}
\end{eqnarray}
which must be satisfied by the open string vertex $V(z)$. In deriving
Eq. (\ref{vertexc}) we have assumed that the generators
of generalized translation annihilate the vacuum: $ {\hat{T}}_i | 0 \rangle =0$.

In the last part of this section we will analyze more explicitly
the case of a constant  background
gauge field satisfying the dipole condition
and living in the identity part  of the $U(N)$ gauge group.
It follows that the first Chern class is given by:
\begin{eqnarray}
\int  {\rm Tr} \left(\frac{q\,F_{12}\mathbb{I}_N}{2\pi} \right)
=  2 \pi \alpha' {q\,F_{12}} N =
n_{12}\equiv f\in \mathbb{Z}\label{fcc1}
\end{eqnarray}
where we have used the fact that $ \int d^2 x = (2 \pi \sqrt{\alpha'})^2$.
In this set-up the cocycle condition given in Eq. (\ref{rc}) can
be easily satisfied by taking:
\begin{eqnarray}
\omega_1 = Q_N &~~~~& \omega_2 = P_N^{f}. \label{gab1}
\end{eqnarray}
In general, the Chan-Paton factor $\Lambda$ depends on the choice of
transition functions.
With the previous choice,
Eq.
(\ref{bst1}) gives the following constraints:
\begin{eqnarray}
 &&e^{2\pi \sqrt{\alpha'}\,i p_1 } \, Q_N\,
{{\Lambda}}_{k} \,Q_N^\dag
 |\Phi, \,k\rangle \equiv {{\Lambda}}_k  |\Phi,\,k\rangle\nonumber \\
 &&e^{2\pi \sqrt{\alpha'}\,i  p_2 }\,
P_N^f\, {{\Lambda}}_k  \, P_N^{-f}
 |\Phi,\,k\rangle \equiv {{\Lambda}}_k  |\Phi,\,k\rangle
\label{bst1a}
\end{eqnarray}
where we have denoted by $k$ the momentum of the physical state
and, as it will be clear later, we have also assumed a dependence of
the Chan-Paton factors on  the momentum $k$.

It is useful to expand the Chan-Paton factors in terms of the
complete set of 't Hooft matrices:
\begin{eqnarray}
( {{\Lambda}}_{k} )_{l m}=\sum_{h_1,\,h_2=0}^{N-1}
C_{(k,\,h_i)}\left( Q_{N}^{h_2} P_{N}^{h_1 }\right)_{l m}.
\label{expa49a}
\end{eqnarray}
We will see in the next subsections  that Eq.s  (\ref{bst1a})
fix the structure of the Chan-Paton factors up to a $c$-number.
When discussing the string vertices, we will explicitly
show that a phase factor is indeed necessary to ensure the correct
hermitian conjugation property of the string vertex.

In the following we are going to treat separately
the degenerate case and the non-degenerate one,
where the great common divisor (GCD) between the first Chern class
and the rank of the gauge group is greater or equal to one, respectively.

\subsection{The non-degenerate case: $GCD(f, N)=1$}
\label{nondege}

In this case, by translating of $2\pi \sqrt{\alpha'}$
the string states  $N$ times along the
$i$-th direction of the torus, Eq.s (\ref{bst1a}) reduce to  the
identity:
\begin{eqnarray*}
 &&e^{2\pi \sqrt{\alpha'}\,i\,N\, p_i } \,
{{\Lambda}}_{k}  \, |\Phi,\,k\rangle
 \equiv {{\Lambda}}_k  |\Phi,\,k\rangle
\label{bst12}
\end{eqnarray*}
where we have used $P_N^N=Q_N^N=1$, giving the following
quantization of the momenta:
\begin{eqnarray}
k_i=\frac{n_i}{\sqrt{\alpha'}\,N}\qquad n_i \in\mathbb{Z}~~
\mbox{and}~~ i=1,\,2 .
\label{mom}
\end{eqnarray}
Eq. (\ref{mom}), when used  in the first of Eq.s (\ref{bst1a}),
yields the following  constraint on the Chan-Paton
factors in Eq. (\ref{expa49a}):
\begin{eqnarray}
{e}^{2 i \pi n_1 /N } Q_{N}   \Lambda_{(n_1,\,n_2 ) } Q_{N}^{-1}
&=&{e}^{2 i \pi n_1 /N }\sum_{h_1,\,h_2=0}^{N-1}
{e}^{ -2 i \pi h_1 /N}  C_{
  (n_1,\,n_2,\,h_1,\,h_2)}Q_N^{h_2}\,P_N^{h_1}
\nonumber\\
&\equiv& \sum_{h_1,\,h_2=0}^{N-1}
C_{ (n_1,\,n_2,\,h_1,\,h_2)}Q_N^{h_2}\,P_N^{h_1}
\label{qw}
\end{eqnarray}
after having used Eq. (\ref{PQ=}) $h_1$ times in the first equality.  Eq. (\ref{qw})
implies
\begin{eqnarray*}
 C_{(n_1,n_2,h_1,h_2)}
= \delta^{[N]}_{n_1, h_1}  C_{(n_1,n_2)} .
\label{Cx1}
\end{eqnarray*}
By using instead Eq. (\ref{bst1a}) along the direction $x_2$ one gets:
\begin{eqnarray}
{e}^{2 i \pi n_2 /N } P_{N}^f   \Lambda_{(n_1,n_2 ) }
P_{N}^{-f} &=&{e}^{2 i \pi n_2 /N }\sum_{h_1,\,h_2=0}^{N-1}
{e}^{ 2 i \pi \,f\, h_2 /N} C_{(n_1,n_2, h_1,\,h_2)}Q_N^{h_2}
\,P_N^{h_1} \nonumber\\
&\equiv& \sum_{h_1,\,h_2=0}^{N-1}C_{(n_1,n_2, h_1,\,h_2)}Q_N^{h_2}\,P_N^{h_1}
\label{pw}
\end{eqnarray}
which implies  that all the quantities $C_{(n_i,h_i)}$ vanish unless
\begin{eqnarray}
f h_2 \equiv -n_2 ~~~~mod~N .
\label{k2h2}
\end{eqnarray}
But, since $Q_N^N=1$, $h_2$ is actually defined modulo $N$.
Hence we can solve Eq. (\ref{k2h2}) as
\begin{eqnarray*}
h_2 &=& \hat h_2 n_2
\label{h2n2}
\end{eqnarray*}
where we have defined the constant $\hat h_2$ as
\begin{eqnarray}
f \hat h_2 \equiv -1 ~~mod~N ~~~~~ 0\le \hat h_2 < N
\label{fhath2}
\end{eqnarray}
in such a way that for any $n_2$ there is only one value of $h_2$ (modulo
$N$). In conclusion,  the periodicity in
$x^1$ and $x^2$ implies for  $C$ the following form:
\begin{eqnarray}
 C_{(n_1,n_2,h_1,h_2)}=
\delta^{[N]}_{h_1, n_1} \delta^{[N]}_{h_2, \hat h_2 n_2}  C _{(n_1,n_2)}
\label{restriC}
\end{eqnarray}
Furthermore,
Eq. (\ref{restriC}) implies
that, for any value of
$n_1$ and $n_2$, we have just
one value of $h_1$ and  one of $\hat{h}_2$ modulo $N$
that allow to satisfy the periodicity in the two directions $x^1$ and $x^2$.
This means that, for each value of the two integers $(n_1 , n_2 )$ we
have a definite value for the integers $(h_1 , h_2)$ modulo $N$  and
this selects
therefore a unique  matrix $(Q_{N}^{h_2} P_{N}^{h_1})_{ab}$ in  the
expansion in Eq. (\ref{expa49a}).
Thus the Chan Paton factors explicitly depend on the momentum.

\subsection{The degenerate case: $g=GCD( f, N)>1$}
\label{dege}

In this case we have that the periodicity conditions
$\omega^N_i=1$ are modified as follows:
\begin{eqnarray*}
(\omega_{1} )^{N} = (\omega_{2})^{\frac{N}{g}} = 1
\label{wra34b}
\end{eqnarray*}
and therefore the momenta become:
\begin{eqnarray*}
k_1=\frac{n_1}{\sqrt{\alpha'}N} \qquad k_2=\frac{n_2 g}{\sqrt{\alpha'}N} . \label{impu}
\end{eqnarray*}
By repeating the same procedure as in the non-degenerate case, we have
that the condition written in Eq. (\ref{qw}) is unchanged, while
Eq. (\ref{pw}) is modified as follows:
\begin{eqnarray*}
{e}^{2 i \pi n_2 g/N} \,P_{N}^{f }
\Lambda_{(n_1,n_2 ) } P_{W}^{-f} & =  &{e}^{2 i \pi n_2 g/N}
\sum_{h_1,h_2=0}^{N-1} C_{(n_1 , n_2 ,h_1 , h_2)}
{e}^{ 2 i \pi f h_2 /N} Q_N^{h_2}\,P_N^{h_1} \nonumber\\
&\equiv &\sum_{h_1,h_2=0}^{N-1} C_{(n_1 , n_2 , h_1 , h_2 ,h_i)}
Q_N^{h_2}\,P_N^{h_1} .
\label{pw1}
\end{eqnarray*}
The following condition is implied:
\begin{equation}
 \frac{ f}{g} h_2 \equiv - n_2~~~~ mod~\frac{N}{g} .
\label{k2h2G}
\end{equation}
The solution of this equation can be found by writing again:
\begin{eqnarray}
h_2=\hat{h}_2\,{n}_2 \qquad {\rm with} ~0\le \hat h_2< N/g
\label{cond1}
\end{eqnarray}
and solving the following condition which is independent on
${n}_2$:
\begin{eqnarray*}
\frac{f}{g} \hat h_2 \equiv -1~~~~mod~\frac{N}{g} .
\end{eqnarray*}
Finally, we can write:
\begin{eqnarray}
\Lambda_{(n_1,n_2 )}\equiv C_{(n_1,\,n_2,n_1,\hat{h}_2\,n_2 )}
Q_N^{ \hat{h}_2\,n_2+m \frac{N}{g}}\, P_N^{n_1} \label{ucsol}
\end{eqnarray}
with $m\in\mathbb{Z}$. However, Eq. (\ref{ucsol}) immediately shows that we are loosing some
solutions. This is  because, due to the periodicity of the matrix
$Q_N$ which is $N$ and not $N/g$, in varying $m$ in the interval
$0\leq m<g$ we have
inequivalent solutions associated to the same momentum $
n_2/\sqrt{\alpha'}N$. This is an extra degeneracy, not present in
the non-degenerate case, which leads us to write, instead of Eq.s
(\ref{cond1}) and (\ref{ucsol}), the following most general
solutions:
\begin{eqnarray*}
h_2 = \hat h_{2}   n_2+ A \frac{N}{g} ~~~~~ 0\le A < g
\label{h2A}
\end{eqnarray*}
and
\begin{eqnarray*}
C_{(n_1,\hat n_2,h_1,h_2)}= \delta^{[N]}_{h_1, n_1}
\delta^{[N]}_{h_2, \hat h_2 \frac{n_2}{g} + A N/g }
C_{(n_1,\,n_2,\,A)}~~~;~~~ 0\le A < g .
\label{CCC}
\end{eqnarray*}
Hence, given $n_1$, we have only one value of $h_1 = n_1$
contributing in the expansion in Eq. (\ref{expa49a}), while, given $n_2$, we
have $g$ possible values  of $h_2$. This means that each value of momentum
has a degeneracy $g$.

In the last part of this section we make some comments on the
generalization of  the previous bundle construction to the case of
magnetized branes living on a product of $T^2 \times T^2 \dots \times T^2$
of  $\frac{\hat{d}}{2}$ tori. The gauge bundle  is again $U(N)$, but now it  is
broken into the product $\prod_{l=1}^{\hat d/2} U(N_l)$ by the
presence, in each factorized torus, of a background gauge field
with constant field strength $ F_{12}^{(l)}$, with $l=1,\dots ,\hat d/2$.
The cocycle conditions, given in Eq. (\ref{rc}), can indeed be
satisfied by embedding, as in the $T^2$ case, the background gauge
fields in the abelian parts of the gauge transition functions
and choosing  for the non-abelian part the product of $\hat d/2$
constant matrices, all equal to the ones given   in Eq. (\ref{gab1}).
This choice of gauge bundle allows  us to trivially generalize the
previous analysis to the product of $T^2 \times T^2 \dots  T^2$
 by simply adopting in
each $T^2$ the procedure developed before. An interesting open
question  is to understand how general  this choice is. It seems
indeed to be consistent when the second Chern numbers are integers
\cite{Hooft81}.  However  it would be nice to find general rules in
constructing consistent gauge bundles associated  to branes
compactified on a generic torus $T^{{\hat{d}}}$. We leave this
analysis to further developments.

\subsection{Boundary state of a non-abelian brane}
\label{boundary}
In this subsection we  derive the boundary state of a space-filling
D25 brane of the bosonic string theory,
whose worldvolume spatial dimensions are partially or
totally compactified on a torus $T^{\hat{d}}$.
The part of the boundary state containing the non-zero modes is the
same as in the uncompactified case. Therefore we need only to
determine the part with the zero modes.
\footnote{In the final writing of this paper we were informed
by D. Duo, R. Russo and S. Sciuto that they have obtained a similar
expression for the boundary state~\cite{duo}.}.
Our results can also be easily extended to the case of  the
D9 brane of a superstring, because the zero
mode structure is the same in the two cases.

The starting point is the computation of the annulus diagram
in the open string channel  and this will allow us to make
contact with previous
results~\cite{Pesando:2005df, Bianchi:2005sa, DiVecchia:2006gg}.
In this calculation the role played by the bundle at string level,
developed in the previous subsections, will be very important.
Then, by using  open/closed string
duality,  we rewrite it in the closed channel and from it we derive
the boundary state.   Subsequently, the
same boundary will be determined directly in the
closed string channel along the lines of Ref. \cite{CLNY88}.
The two approaches provide the same boundary state up to
a phase that, however, does not contribute to the annulus diagram.

In order to use all the machinery  developed in the previous
sections we restrict ourselves to the background $\mathbb{R}^{1,25-\hat{d}}\times (T^{2})^{\hat{d}/2}$
($\hat{d}$ even).
This simplification, however,  will not be  necessary for determining
the boundary state directly from the closed string channel.
On each of the tori  we turn on a background gauge field
with constant field strength given by:
\begin{eqnarray*}
F_l\equiv \left(\begin{array}{cc}
            0       &F_{12}^l\\
             -F_{12}^l  &0
            \end{array}\right)
            \,\mathbb{I}_{N_l} .
\label{F12}
\end{eqnarray*}
Let us start by computing the annulus diagram in the open string
channel in the case in which the open strings are attached to two
space-filling branes having the same gauge field on their worldvolume
and therefore
satisfying the dipole condition. This amounts to evaluate:
\begin{eqnarray}
{\cal Z}^{dipole}_{25;\,F}= M^2 \int_0^\infty \frac{d \tau}{\tau}{\rm Tr}\left[
 e^{-2\pi \tau L_0}\right]
\label{freen}
\end{eqnarray}
where, in order to be more general, we have considered $M$
space-filling branes
producing the factor $M^2$ in the previous equation.
These are degrees
of freedom associated  to an additional
$U(M)$ gauge group under which the background gauge
fields are uncharged.  $L_0$ is the open string
Hamiltonian given in Eq. (\ref{L0ni62}).
After some calculation, Eq. (\ref{freen}) in the
non-degenerate case becomes:
\begin{eqnarray}
{\cal Z}_{25;\, F}^{dipole}&=& \frac{M^2
V_{26-\hat d}}{(8\pi^2\alpha')^{(26-\hat d)/2}} \,
\int_{0}^{\infty}  \frac{d\tau}{\tau} \tau^{-\frac{26-\hat d}{2}}
\left[f_1(e^{-\pi\tau})\right]^{-24}  \nonumber\\
&\times&\prod_{l=1}^{\hat d/2} \left[ \sum_{n^{(l)}\in
\mathbb{Z}}e^{-2\pi\tau
\sum_{p,q=1}^2\left[\frac{n_p^{(l)} }{N_l}( {\cal G}^{(l)} )^{pq}
\frac{n _q^{(l)} }{N_i}\right]}\right].
\label{z25a}
\end{eqnarray}
This equation can be rewritten in the closed string channel by using the
following Poisson resummation formula:
\begin{eqnarray*}
\sum_{w\in\mathbb{Z}^p} {\rm
exp}\left[-\pi(w+x) A(w+x)\right] =
\left(\det A\right)^{-1/2}\sum_{w\in\mathbb{Z}^p} {\rm
exp}\left[-\pi\,w  A^{-1} w + 2\pi i\,w \,x\right]
\label{ffun}
\end{eqnarray*}
and the transformation properties of the function $f_1 (e^{- \pi \tau}) =
\sqrt{t}f_1 (e^{- \pi t}) $ under $\tau \rightarrow t =
\frac{1}{\tau}$,
obtaining:
\begin{eqnarray}
{\cal Z}_{25;\, F}^{dipole}&=& \frac{M^2
V_{26-\hat d}}{(8\pi^2\alpha')^{(26-\hat d)/2}}
\,\prod_{l=1}^{\hat{d}/2} \left[ \left({\rm det}\frac{{\cal
G}_{pq}^{(l)}}{2}\right)^{1/2}\,N^2_l \right]\nonumber\\
&\times& \int_{0}^{\infty} d t~ \left[f_1(e^{-\pi t})\right]^{-24}~
 \prod_{l=1}^{\hat{d}/2} \left[\sum_{s_{(l)} \in \mathbb{Z}}
 e^{-\frac{\pi}{2} t s^p_{(l)} N_l
 {\cal G}^{(l)}_{pq}
 s^q_{(l)} N_l}\right].
\label{bicsa}
\end{eqnarray}
Eq. (\ref{bicsa}) gives the interaction between two stacks of $M$
D25 magnetized branes in the closed channel. By using the equation:
\begin{eqnarray*}
{\cal Z}_{25;\, F}^{dipole} = \langle D25(E,\,F) | \Delta |D25(E,\,F)   \rangle
\label{clobou8}
\end{eqnarray*}
where $\Delta$ is the closed string propagator, we can determine
the boundary state apart from an overall  phase.

In this way  we get the following expression of the boundary state, which we now write both
for the D25 brane of the bosonic theory and for the
D9 brane of the superstring:
\begin{eqnarray}
&&\!\!\!\!\!| D(d-1)(E,\,F) \rangle  =
\frac{ \sqrt{ {\rm det}\, {\cal E}} }{({\rm det} G)^{1/4}}
 M  \frac{T_{d-1}}{2}
 e^{-\sum_{n=1}^\infty \frac{1}{n} ~\alpha^{\mu}_{-n} G_{\mu \nu}
\alphat^{\nu}_{-n} }  \nonumber \\
&&\!\!\!\!\!\times ~e^{-\sum_{n=1}^\infty \frac{1}{n} ~\alpha^{i}_{-n}
\calE_{ik} (\calE^{-T})^{kh}G_{hj}   \alphat^{j}_{-n} } ~| k=0; 0_a,
0_{\tilde a} \rangle \prod_{l=1}^{\hat d/2}| D(d-1)(E,\,F)^l_{N_l} \rangle_{z.m.}.
\label{Bzemo7}
\end{eqnarray}
Here, $d=26$ or $10$, $T_{d-1}$ is the tension of
the space-filling brane given just before Eq. (\ref{plain-D25a}), $\mu, \nu$ are the non-compact space-time indices  and
\begin{eqnarray}
|D(d-1)(E,\,F) ^l_{N_l} \rangle_{z.m.} =
N_l \sum_{n^{(l)} , m^{(l)}\in \mathbb{Z}}
\delta_{\hat{n}_p^{(l)} - 2 \pi \alpha ' q F_{pq}^{l}
  \hat{m}_{q}^{(l)}} | n_{p}^{(l)}, m_{q}^{(l)} \rangle
\label{bou73a}
\end{eqnarray}
where ${\cal{E}}$ is given in Eq. (\ref{calE}). Notice that
the particular structure of the delta  function is the one that is also
required  by the overlap conditions in Eq. (\ref{zeromo}).
Finally, in order to reproduce the peculiar structure of the exponent in Eq. (\ref{bicsa}) and also
to implement that the first Chern-class is integer as
dictated by Eq. (\ref{fcc1}),  we have
to impose that $m^q_{(l)}=N_l s^q_{(l)}$ with $s^q_{(l)}\in \mathbb{Z}$
and we end with the following boundary state:
\begin{eqnarray}
|D(d-1)(E,\,F) ^l_{N_l}\rangle_{z.m.} = N_l \sum_{ s_{(l)}\in \mathbb{Z}}
|2\pi\alpha' N_l\,qF^l_{pq}
\,s^q_{(l)}, N_l\,s^q_{(l)}\rangle
\label{bou73ba}
\end{eqnarray}
Before discussing the  degenerate case let us
compare the boundary state in Eq. (\ref{bou73ba})
with the one we have exhibited
in Eq. (17) of Ref.~\cite{DiVecchia:2006gg}.
They only differ from the fact that
the zero modes in Eq.s (\ref{bou73a}) and (\ref{bou73ba}) have  integer
Kaluza-Klein momenta and winding numbers, while
the ones in Eq. (17) of Ref.~\cite{DiVecchia:2006gg} have
integer winding modes but fractional Kaluza-Klein
modes.  We think, however, that it is unnatural to have fractional
Kaluza-Klein  momenta in the closed string sector and therefore we
prefer the boundary state given here which eliminates this feature.

The previous equations can be easily generalized to the
degenerate case. Eq. (\ref{z25a}) now  becomes
\begin{eqnarray}
{\cal Z}_{25;\, F}^{dipole}&=& \frac{M^2
V_{26-\hat d}}{(8\pi^2\alpha')^{(26-\hat d)/2}} \,\left[\prod_{l=1}^{\hat d/2}
g^{(l)}\right] \,
\int_{0}^{\infty} \frac{d\tau}{\tau} \tau^{-\frac{26-\hat d}{2}}
\left[f_1(e^{-\pi\tau})\right]^{-24}\nonumber\\
&\times&\prod_{i=l}^{\hat d/2} \left[ \sum_{n^{(l)}\in
\mathbb{Z}}e^{-2\pi\tau
\sum_{p,q=1}^2\left[\frac{n_p^{(l)}(\delta^p_1+g^{(l)}\,\delta^p_2
)}{N_l}({\cal G}^{(l)})^{pq}\frac{n _q^{(l)} (\delta^q_1+g^{(l)}
\,\delta^q_2) }{N_i}\right]}\right]
\label{z25}
\end{eqnarray}
where  the overall
factor $\left[\prod_{l=1}^{\hat d/2} g^{(l)}\right]$ and the peculiar
structure of the momenta are due respectively to the degeneracy and the
structure of the dipole string momentum  in the degenerate case.

By rewriting it in the closed string channel we get:
\begin{eqnarray}
{\cal Z}_{25;\, F}^{dipole}&=&
\frac{M^2
V_{26-\hat d}}{(8\pi^2\alpha')^{(26-\hat d)/2}}
\prod_{l=1}^{\hat{d}/2}\left[\left({\rm det}\frac{{\cal
G}_{pq}^{(l)}}{2}\right)^{1/2}\,N^2_l\right]
 \, \int_0^\infty  d t~\left[f_1(e^{-\pi t})\right]^{-24}~        \nonumber\\
 &\times& \prod_{l=1}^{\hat{d}/2} \left[\sum_{s_{(l)}\in \mathbb{Z}}
 e^{-\frac{\pi}{2} t s^p_{(l)} N_l \left( \delta^p_1+
 \frac{\delta^p_2}{ g^{(l)}} \right) {\cal G}^{(l)}_{pq} \left( \delta^q_1+
 \frac{\delta^q_2}{ g^{(l)}} \right) s^q_{(l)} N_l }\right].
\label{bics}
\end{eqnarray}
The zero-mode structure of the boundary state that is
extracted from the previous
equation has again the form given in Eq.s (\ref{bou73a}) and (\ref{bou73ba}),
but now one can impose a weaker condition:
\begin{eqnarray*}
(2\pi\alpha') \frac{N_l}{g^{(l)}}qF^l_{12}=\frac{f^l}{g^{(l)}}\in
\mathbb{Z}
\label{weak54}
\end{eqnarray*}
which requires that $m^q_{(l)}= N_l/g^{(l)}\,s^q_{(l)}$. However,  in order to
reproduce from the boundary state  the expression in Eq. (\ref{bics})
for  the degenerate case, we need to take
$m^q_{(l)}=N_l\,s^q_{(l)}\,(\delta_q^1+\delta_q^2/g^{(l)})$. By collecting all
the results we can write:
\begin{eqnarray}
| D25(E,\,F)^l_{N_l}\rangle_{z.m.} = N_l \sum_{ s_{(l)}\in \mathbb{Z}}
|2\pi\alpha' N_l\,qF^l_{pq}
\,s^q_{(l)}\,(\delta_q^1+\delta_q^2/g^{(l)}),
N_l\,s_{(l)}\,(\delta_q^1+\delta_q^2/g^{(l)})\rangle
\label{bou73b}
\end{eqnarray}
The generalization of the previous expression
to the $D9$ brane is straightforward.
It is easy  to verify that the boundary state in Eq. (\ref{bou73b}) reproduces
the zero-mode contribution in  Eq. (\ref{bics}).
Notice that the asymmetry between directions $1$ and $2$
is a direct consequence of the asymmetric
choice for the transition function performed in Eq. (\ref{gab1}).

In the following we would like to explore the
connection  between the magnetized D25 branes carrying
non-trivial gauge bundles and the T-dual systems  corresponding
to lower dimensional branes generically wrapping  some cycles of the torus.
The T-duality which we consider is the one that exchanges the
K\"ahler structures $T_i^{(l)}$ with the complex structures
$U_i^{(l)}$ of the torus $ T^2$ defined as follows:
\begin{eqnarray*}
T^{(l)} \equiv T_1^{(l)}   + i T_2^{(l)} =
B_{12}^{(l)} + i \textsc{}\sqrt{G^{(l)}}~~;~~
U^{(l)} \equiv U_1^{(l)} + i  U_2^{(l)} =
 \frac{G_{12}^{(l)}}{G_{11}^{(l)}}  + i \frac{\sqrt{G^{(l)}}}{G_{11}^{(l)}}.
\label{tau78}
\end{eqnarray*}
The exponential factor in Eq. (\ref{z25}), which is essentially due to
the zero modes of the open
string Hamiltonian~\footnote{In the non-degenerate
case and in the case of a squared torus this Hamiltonian appears in
Ref.s ~\cite{9810072,Blumenhagen:2000fp,Blumenhagen:2000wh}}
on the torus $T^2$,
can be written as:
\begin{eqnarray}
\left( \frac{n_p^{(l)}\delta_1^p+n_p^{(l)}g^{(l)}\delta_2^p}{N_l}
\right)\, ({\cal
G}^{(l)})^{pq}\,\left(\frac{n_q^{(l)}\delta_1^q+n_q^{(l)}g^{(l)}
\delta_2^q }{N_l}\right) =
\frac{T_2^{(l)}}{U_2^{(l)}} \frac{|n_2^{(l)}-n_1^{(l)}
\frac{U^{(l)}}{g^{(l)}}|^2}{|\frac{N_l}{g^{(l)}}T-
\frac{f^l}{g^{(l)}}|^2}\label{Hmos}
\end{eqnarray}
where we have used the open string metric on the torus $T^2$:
\begin{eqnarray*}
({\cal{G}}^{(l)})^{pq}  =
\frac{T_2^{(l)}}{ U_2^{(l)} ( {T_{2}^{(l)}}^{2} + {{\cal{B}}^{(l)}}^{2}) }
\left( \begin{array}{cc} |U^{(l)}|^2 & - U_1^{(l)} \\
                       - U_1^{(l)}  & 1 \end{array} \right)~~;~~{\cal{B}} =
                       B_{12} - 2 \pi \alpha' q F_{12} .
\label{opeme}
\end{eqnarray*}
Under the T-duality transformation, i.e.  $T \leftrightarrow U$,
the l.h.s. of Eq.  (\ref{Hmos})   becomes:
\begin{eqnarray}
\frac{T_2^{(l)}}{U_2^{(l)}} \frac{|n_2^{(l)}-n_1^{(l)}
\frac{U^{(l)}}{g^{(l)}}|^2}{|\frac{N_l}{g^{(l)}}T-\frac{f^l}{g^{(l)}}|^2}
 \Longrightarrow \frac{U_2^{(l)}}{T_2^{(l)}}
 \frac{|n_2^{(l)}-(w_1^{(l)} +
 \frac{v_1^{(l)}}{g^{(l)}})
T^{(l)}|^2}{|\frac{N_l}{g^{(l)}}U-\frac{f^l}{g^{(l)}}|^2}
\label{ibh}
\end{eqnarray}
where we have rewritten $n_1^{(l)}/g^{(l)}$ as $w_1^{(l)} +
v_1^{(l)}/g^{(l)}$ with $w_1^{(l)}\in \mathbb{Z}$ and
$v_1^{(l)}=0,\dots, g^l-1$. The T-dual zero mode Hamiltonian in the
non-degenerate case can be interpreted as the one of a lower dimensional brane
wrapping respectively $(\pm N_l,\,\mp f^l)$ times the two
one-cycles of the torus as we have discussed in the introduction.
This can be seen by comparing the zero-mode Hamiltonian
in the r.h.s. of Eq. (\ref{ibh}) for the squared torus $T^2$ with $B_{12}=0$
$(T=i \frac{R_1 R_2}{\alpha'}; U = i \frac{R_2}{R_1})$,
with the zero-mode Hamiltonian given for instance
in Sect. (3.1) of Ref.~\cite{0503179}.

In the degenerate case we see instead that, for $v_1^{(l)}=0$, the
open string Hamiltonian coincides with the one of a lower-dimensional brane with wrappings
$n_l=\pm N_l/g^{(l)},\,m_l=\mp f^l/g^{(l)}$  along the one-cycles of the torus.
For $v_1^{(l)}\neq 0$, Eq. (\ref{ibh}) shows that
also for zero winding $w_1^l=0$ the open  string has a minimal
length and therefore the previous Hamiltonian describes the
interaction between parallel branes displaced in the space transverse to
their worldvolume~\footnote{See for
instance sect. (3.2) of Ref.~\cite{0503179}.}.
In this case, the zero-mode contribution to the $D$-brane interaction, in the
T-dual configuration, can be written as:
\begin{eqnarray*}
{\cal Z}_{\rm z.m.}= \prod_{l=1}^{\hat{d}/2} \left[g^{(l)} \sum_{(k^{(l)},\,
w^{(l)})\in \mathbb{Z}}\sum_{v^{(l)}=0}^{g^{(l)}-1} e^{-2\pi\tau
\frac{U_2^{(l)}}{T_2^{(l)}} \frac{|k^{(l)}-(w^{(l)}
+\frac{v^{(l)}}{g^{(l)}}) T^{(l)}|^2}{|n_l\,U+m_l|^2}}\right] .
\end{eqnarray*}
where the sum over $v^{(l)}$
suggests that the brane is not stable and decays into
a stack of $g^{(l)}$ branes wrapped
$m_l \equiv \mp f^l/g^{(l)}$ and $ n_l \equiv \pm
N_l/g^{(l)}$  times along the cycles of the
tori $(T^{2})^{(l)}$~\footnote{See
the discussion on page 84 of Ref.~\cite{0610327}.}.

In the last part of this subsection we derive   the boundary
state with a gauge field on it directly in the closed string channel
starting from the boundary state without a gauge field and
following the procedure described in Ref.~\cite{CLNY88} that provides
the following expression:
\begin{equation*}
|D25(E,F)\rangle
= Tr\left( P~e^{-i\oint  q A} \right) ~
|D25(E,F=0)\rangle
\label{bsf1}
\end{equation*}
where the boundary state without the gauge field is given in Eq.s
(\ref{plain-D25a}) and (\ref{D25zmta}). The previous
path ordering is explicitly evaluated in Appendix \ref{bouclosed}
getting
\begin{eqnarray}
|D25(E,F)\rangle
&=&
\frac{T_{25}}{2}
~N
\frac{\sqrt{\det \calE}}{( \det G)^{1/4}}
\sum_s
e^{-i \pi \hat F^<_{i j} s^i s^j}
|n_i= \hat F_{i j}N s^j, m^i=s^i\rangle
\nonumber\\
&\times& e^{-\sum_{n=1}^\infty \frac{1}{n} ~\alpha^{i}_{-n}
\calE_{ik} (\calE^{-T})^{kh}G_{hj}   \alphat^{j}_{-n} } ~| 0_a,
0_{\tilde a} \rangle
\nonumber\\
&\times& e^{-\sum_{n=1}^\infty \frac{1}{n} ~\alpha^{\mu}_{-n} G_{\mu \nu}
\alphat^{\nu}_{-n} } ~| k=0 \rangle.
\label{D25ph}
\end{eqnarray}
Here $\hat F^<_{i j}={\hat{F}}^{ij}$ when $i<j$ and zero otherwise.
The boundary state in Eq. (\ref{D25ph}) differs
from the one given in Eq. (\ref{bou73ba}) for a phase factor.
However, this extra phase does not give any contribution to
the one-loop free-energy and this is
the reason why in the previous determination
of the boundary it has not been  possible to reveal its presence.

\subsection{The string vertices}
\label{strive}

In this section we construct the vertex operators  corresponding to
the open strings having their endpoints on the D25  branes. We
limit our analysis
to the compactified part of the vertex and also to
the lowest state, the tachyonic one, being the
generalization to higher state vertices straightforward.

In the non-degenerate case and on the simplest case of $T^2$,
the compact part of the string vertex, describing an open-string tachyon
living on a non-abelian brane is
given by:
\begin{eqnarray*}
V(x;k)= {e}^{ i k_i  X^i (x) }
 \Lambda_{(k_1 , k_2)}
 \label{svt2ata}
\end{eqnarray*}
where $ X^i (x)$ is given in Eq. (\ref{icso}) with $\sigma =0$ and $x
= e^{-i \tau}$, $\Lambda$ is the Chan-Paton factor and
the momentum is given by
\begin{eqnarray*}
(k_1,k_2) = \frac{1}{\sqrt{\alpha '}}
(\frac{n_1}{ N  }, \frac{n_2}{N  } ) .
\label{nondebra}
\end{eqnarray*}
Using Eq. (\ref{icsot}) it is easy to rewrite the
previous equation as follows:
\begin{eqnarray}
V(x;k)= {e}^{ i k_i  X_{L(0)}^i(x) }
 \Lambda_{(k_1 , k_2)}
 \label{svt2a}
\end{eqnarray}
where we have neglected cocycle factors, an example of which is provided for instance by the last term
in Eq. (\ref{cofa}).
The generalization of such a vertex to the
case of $(T^2)^{\hat{d/2}}$ is simply
the factorized product of $\frac{\hat{d}}{2}$ operators, one for each torus
$T^2$, with the same structure as the one given in Eq.
(\ref{svt2a}).

In sections (\ref{nondege}) and (\ref{dege}) we have determined the
structure of the Chan-Paton factors, respectively in
the non-degenerate and degenerate case, up to a c-number
factor. In particular, for the non-degenerate case, the Chan-Paton
factor is given in Eq. (\ref{expa49a}) together with Eq. (\ref{restriC}).
Since now the Chan-Paton factor depends on the momentum, we must also
remember that the vertex operator in Eq. (\ref{svt2a}) has to   satisfy
the hermitian conjugation property
\begin{eqnarray*}
V(z;k)^\dag= \frac{1}{{z}^{2h} }V(1/{z};-k)
\label{cos1}
\end{eqnarray*}
with $h$ being the conformal weight,
which  imposes the following
constraint on the Chan-Paton factor:
\begin{eqnarray}
\Lambda_{(k_1 , k_2)}^\dagger =\Lambda_{(-k_1 , -k_2)} .
\label{Lambda-herm}
\end{eqnarray}
In order to satisfy the previous identity we must add a phase factor
to the Chan-Paton factor determined above and we get:
\begin{eqnarray}
\Lambda_{(k_1 , k_2)}=
\frac{1}{\sqrt{N}} ~e^{i \pi
N\,\alpha' \, \hat h_2 k_1 k_2 } \left(
Q_{N}^{N\,\sqrt{\alpha'}\,\hat h_2 k_2} P_{N}^{
N\,\sqrt{\alpha'} k_1 }\right)
\label{cpfndc}
\end{eqnarray}
where ${\hat{h}}_2$ is defined in Eq. (\ref{fhath2}). The fact that
$\Lambda$ in Eq. (\ref{cpfndc}) satisfies Eq. (\ref{Lambda-herm}) is a
consequence of the relations: $P^{\dagger} = P^{-1}$ and
$Q^{\dagger} = Q^{-1}$.

It is easy to check that the $\Lambda_{(k_1,\,k_2)}$ matrix
satisfies  the multiplication rule
\begin{eqnarray*}
\Lambda_{(k_1 , k_2) a c} \Lambda_{(l_1 , l_2) c b} =
\frac{1}{\sqrt{N}} ~e^{i k \wedge l } ~\Lambda_{(k_1+l_1 ,
k_2+k_2) a b} \label{Lambda-prod}
\end{eqnarray*}
where we have introduced the product
\begin{eqnarray*}
k \wedge l = \pi \alpha' ~N \hat h_2 (k_1 l_2 - k_2 l_1)
\label{NA-wedge}
\end{eqnarray*}
or more in general:
\begin{eqnarray*}
\prod_{i=1}^M \Lambda_{(k_1^{(i)} ,
k_2^{(i)})}= N^{\frac{1-M}{2}}
\prod_{i<j=1}^M e^{i k^{(i)}
\wedge k^{(j)}}\Lambda_{(\sum_{i=1}^M k_1^{(i)}, \,\sum_{i=1}^M k_2^{(i)})} .
\end{eqnarray*}
In order to compare with an alternative description of wrapped branes that we will present in sect. 4, it is useful to evaluate explicitly the trace over the Chan Paton factors, that
 is given by:
\begin{eqnarray}
\mbox{Tr} \left[ \prod_{i=1}^M \Lambda_{(k_1^{(i)}, k_2^{(i)})} \right] &=&
N^{-\frac{M}{2}} \prod_{i < j=1}^{M} e^{i k^{(i)} \wedge k^{(j)} }
e^{i \pi N \alpha' \hat{h}_{2} \sum_{i,j=1}^{M} k_{1}^{(i)} k_{2}^{(j)} }\nonumber\\
&\times&\delta^{[N]}_{N \sqrt{\alpha'} \sum_{i=1}^{M} k_{1}^{(i)};0 }
\delta^{[N]}_{N \sqrt{\alpha'} \sum_{i=1}^{M} k_{2}^{(i)};0 }
\label{trace2}
\end{eqnarray}
The normalization coefficient $\frac{1}{\sqrt{N}}$
introduced in Eq. (\ref{cpfndc}) is there to
ensure that the trace over two Chan-Paton factors is independent on the number of
colors.

The analysis done so far to determine the structure of the  string
vertices in the non-degenerate case can be easily extended to
the degenerate case. One gets again the vertex
\begin{eqnarray*}
V(x;k,A)= {e}^{ i k_i  X_{L(0)}^i(x) }
 \Lambda_{(k_1 , k_2),A}
\end{eqnarray*}
where the momentum is given by:
\begin{eqnarray*}
(k_1,k_2)  =
\frac{1}{\sqrt{\alpha '}} \left(\frac{n_1}{  N  }, \frac{
n_2}{N/g  }\right) .
\label{NA-deg-momenta}
\end{eqnarray*}
By analogy with Eq. (\ref{cpfndc}) we take, for the momentum
dependent Chan-Paton factor, the following
expression:
\begin{equation*}
 \Lambda_{(k_1 , k_2),A}
= \frac{1}{\sqrt{N}} ~e^{i \frac{\pi}{N} n_1 \left(  \hat h_2  n_2 + A
  N/g \right)}
\left( Q_{N}^{\hat h_2 n_2+ A  \frac{N}{g}} P_{N}^{n_1
}\right).
\end{equation*}
It satisfies the hermiticity property
\begin{eqnarray*}
\Lambda_{(k_1 , k_2),A}^\dagger =\Lambda_{(-k_1 , -k_2), -A}
\label{hermpr}
\end{eqnarray*}
and  the multiplication rule
\begin{eqnarray*}
\Lambda_{(k_1 , k_2),A } \Lambda_{(l_1 , l_2),B} =
\frac{1}{\sqrt{N}} ~e^{i \pi \frac{\sqrt{\alpha'}N}{g} (k_1 B- l_1 A)} ~e^{i
\frac{1}{g} k \wedge l } ~\Lambda_{(k_1+l_1 ,
k_2+k_2),A + B } .
\label{Lambda-deg-prod}
\end{eqnarray*}

\section{Narain branes}
\label{sect-narain-closed}

In the previous section we have constructed the boundary state and
the open string vertex operators corresponding to  wrapped
space-filling branes with a background gauge field living on their
worldvolume. They are described by  gauge bundles. However, as we
have pointed out in the introduction, this is not necessarily the
unique way of describing wrapped magnetized space-filling branes and in this
section we discuss another kind of space-filling branes, the {\em
Narain branes}. Their name is due to the fact that they can be
obtained from the usual branes without a background gauge field by
means of a transformation of the Narain T-duality group
$O(\hat{d},\hat{d},\Z)$ which is reviewed in Appendix
\ref{T-duality}. We construct the boundary state corresponding to
this kind of branes and show that it is coincident (up to a phase
which does not contribute to the one-loop vacuum amplitude) with the
one already constructed in the previous section for the gauge
bundles. Then we add Wilson lines to this boundary state in the case $F=0$ in order
to describe a $D$-brane wrapped $N$ times around a torus and analyze their transformation properties
under the Narain group,
and then generalize to the case $F\neq 0$.
We give the vertex operators for the open strings having their
endpoints attached to the Narain branes, showing that their
scattering amplitudes with closed strings are different from those
that one derives from the gauge bundles. In all the examples we will
explicitly refer to the tachyon vertex because it encodes the main
features of the problem, the generalization to all other vertices
being straightforward.

\subsection{Narain branes from plain brane: general case.}
\label{nabrplbr}

In this section we consider the bosonic string, taking as our starting point a  D25 brane
in a generic constant closed string background $E^t$ with
no background gauge field $(F^{t} =0)$ on its worldvolume.
By applying on it a general
transformation of the T-duality group, we  get what we call the most
general  Narain brane having  a non-vanishing constant magnetic field $F$
on its worldvolume.

We start with a plain  $D25$ on $R^1\otimes T^{25}$ whose boundary state
satisfies  the boundary conditions in Eq. (\ref{bou0bs}) with $F^t =0$:
\begin{eqnarray}
\left[G_{ij}^t{(\dot X^t)}^j+ B_{ij}^t
{({X'}^t)}^j\right]_{\tau=0}|D25(E^t,F^t =0)\rangle \equiv  { P}^t_i
|D25(E^t,F^t=0)\rangle = 0
\label{D25t}
\end{eqnarray}
The solution of these equations is given in Eq.s (\ref{plain-D25a}) and
(\ref{D25zmta}) for $d=26$.

We now perform a canonical transformation as in
Eqs. (\ref{CanTras}), (\ref{lam}) such that  $\D^{-1} \C$
is a well-defined quantity, then the boundary defining
Eq. (\ref{D25t}) becomes
\begin{eqnarray}
\left[{ P}_i + ( \D^{-1} \C)_{i j}~ \frac{X'^j}{2\pi \alpha'}
\right]|D25(E ,  F\rangle = 0
\label{120}
\end{eqnarray}
that is equal to  Eq. (\ref{bou0bs}) with the following gauge field:
\begin{eqnarray}
\hat F=  2\pi\alpha' q  ~F= -\D^{-1}\C =  \C^T \D^{-T}.
\label{effe}
\end{eqnarray}
The last equality follows from the entry $(2,2)$ of Eq. (\ref{LLT}).
Here $q$ is the electric charge.

Under this transformation
the zero mode part of the boundary
(\ref{D25zmta}) becomes:
\begin{eqnarray}
|D25(E,F)\rangle _{z m}&=& | D25(E^t,F^t=0) \rangle _{z m}
\nonumber\\
&=& \sqrt{\det \D} \frac{\sqrt{\det { \cal E}}}{\left(\det
G\right)^{1/4}}~ \sum_{s\in Z^{25}} | n_i= (\C^T)_{i j} s^j, m^{i}=
{\D^T}{}^i_{ j}s^j\rangle
\nonumber\\
&=& \sqrt{\det \D} \frac{\sqrt{\det { \cal E}}}{\left(\det
G\right)^{1/4}}~ \sum_{s\in Z^{25}} | n_i= 2\pi \alpha' q  F_{i j}
m^j, m^{i}= {\D^T}{}^i_{ j}s^j\rangle
\nonumber\\
\label{D25zm}
\end{eqnarray}
where in the first equality we have written the boundary state with a
non-vanishing gauge field as the one in Eq. (\ref{D25zmta}) with
$n_{i}^{t} =0$ and $m^{t\,\,i} = s^i$. By rewriting those
variables in terms of $n$ and $m$ given in the upper equation in
(\ref{inverse-mn-a}) one gets the second line of Eq. (\ref{D25zm})
that can finally be written as in the third line
by means of Eq. (\ref{effe}). A detailed explanation of the
normalization factor is given in Ref.~\cite{DiVecchia:2006gg} and reviewed in Appendix~\ref{T-duality}.
Finally, when $\det \D\ne0$, the complete boundary state (\ref{plain-D25a})
satisfying Eq. (\ref{120}) is:
\begin{eqnarray}
 |D25(E,F)\rangle
&=& \frac{T_{25}}{2} \sqrt{\det \D} \frac{\sqrt{\det { \cal
E}}}{\left(\det G\right)^{1/4}}~ \sum_{s\in Z^{25}} | n_i= 2\pi
\alpha' q  F_{i j} m^j, m^{i}= {\D^T}{}^i_{ j}s^j\rangle
\nonumber\\
&\times& e^{-\sum_{n=1}^\infty \frac{1}{n} ~\alpha^{i}_{-n}
\calE_{ik} (\calE^{-T})^{kh}G_{hj}   \alphat^{j}_{-n} } ~| 0_a,
0_{\tilde a} \rangle
\nonumber\\
&\times& e^{-\sum_{n=1}^\infty \frac{1}{n} ~\alpha^{0}_{-n} G_{0 0}
\alphat^{0}_{-n} } ~| k_0=0 \rangle. \label{D25}
\end{eqnarray}
By construction this boundary state satisfies Eq. (\ref{bou0bs})
with ${\hat{F}}$ given in Eq. (\ref{effe}). Notice also that this construction
is valid for an arbitrary torus $T^{{\hat{d}}}$.

\subsection{Special cases}
\label{T-dual-fact-torus-B}

{\em T-duality  on a factorized torus}

\noindent

In Sect. \ref{sect-nonabelian-open}  we have given the boundary state of a wrapped magnetized brane in the gauge bundle description.
We would like here to compare this with the boundary state corresponding to a Narain brane.
To this aim,
we consider the simple
case in which  the compact space
is a factorized torus.
In particular, we can focus on a single $T^2$, being
the generalization to $(T^2)^{\hat d/2}$
straightforward, and as a very special example we consider
the canonical
transformation acting in the first torus $T_{(1)}^2$ along
the directions $1$ and $2$
realized by the matrix:
  \begin{eqnarray}
  \Lambda_2(p_{(1)},\,q_{(1)})=\left(\begin{array}{cc}
                          r_{(1)}\mathbb{I} & is_{(1)}\,\sigma_2\\
                          -ip_{(1)}\,\sigma_2&q_{(1)}\mathbb{I}
                          \end{array}\right) .
\label{Lambda2}
  \end{eqnarray}
  By imposing the condition $\Lambda_2(p_{(1)},q_{(1)}) \in O(2,2,Z)$ (see Eq. (\ref{CanTransMatrix})) we
  get:
  \begin{eqnarray*}
  J=\left(\begin{array}{cc}
            0&(r_{(1)}\,q_{(1)}-s_{(1)}\,p_{(1)})\mathbb{I} \\
             (-s_{(1)}\,p_{(1)}+r_{(1)}\,q_{(1)})\mathbb{I}&0
                          \end{array}\right)
  \end{eqnarray*}
  which implies that $r_{(1)} q_{(1)}- p_{(1)} s_{(1)} =1 $. From Eq.s (\ref{lam})
  and (\ref{Lambda2}) we can see that:
  \begin{eqnarray*}
  \det \D = q_{(1)}^2.
  \label{detD}
  \end{eqnarray*}
  This T-duality transforms a plain $D$-brane into a configuration of a $D$-brane with a gauge field strength
  given by (see Eq. (\ref{effe})):
\begin{eqnarray*}
 2\pi q \alpha' F=- \D^{-1}\C=\left(\begin{array}{cc}
          0&\frac{p_{(1)}}{q_{(1)}} \\
           -\frac{p_{(1)}}{q_{(1)}}&0
           \end{array}\right)=\frac{p_{(1)}}{q_{(1)}}i\sigma_2.
\label{effe43}
\end{eqnarray*}
{From} this equation we can see that $ 2\pi q \alpha' F_{12}$ is an integer number. This realizes Eq. (\ref{chernc89a}) according to the second logical possibility discussed in the Introduction, not coming the integer $q_{(1)}$ from any trace over the gauge group. The latter condition is the first hint that the Narain branes
are branes wrapped $q_{(1)}$ times on the entire torus.

In this case Eq. (\ref{120}) becomes
\begin{eqnarray*}
[{ P}_1 - \frac{p_{(1)}}{q_{(1)}} \frac{ X'^2}{2 \pi \alpha'}] | D25(E,F)
\rangle = [{ P}_2 + \frac{p_{(1)}}{q_{(1)}} \frac{X'^1}{2 \pi \alpha'}] |
D25(E,F) \rangle = 0 \label{ptra}
\end{eqnarray*}
where the zero-mode part of the boundary is now given by (see Eq.
  (\ref{D25zm}))
  \begin{eqnarray}
 \!\!\!\! \!\!\!\!\!\!\!\!|D25(E,F)\rangle _{z m}\!\! &=& \!\!\! \frac{q_{(1)}
\sqrt{\det {\cal{E}}_{(1)}}}{
  \left(\det G_{(1)}\right)^{1/4}}
 \!\! \sum_{s^1,s^2\in Z}\!\! | n_1= p_{(1)} s^2,n_2=- p_{(1)} s^1,
  m^{1,2}= q_{(1)} s^{1,2}\rangle
\label{D25zm-fact-1}
\end{eqnarray}

The factor $q_{(1)}$ appearing in front of the boundary in the previous equation  confirms that a Narain brane can be interpreted as a brane wrapping $q_{(1)}$ times the whole torus. Indeed the area of such an object, its Dirac-Born-Infeld action (DBI) and therefore the boundary state normalization should be proportional to $q_{(1)}$ because the brane covers $q_{(1)}$ times the compact manifold and this indeed happens
in Eq. (\ref{D25zm-fact-1}).

The boundary state in Eq. (\ref{D25zm-fact-1}) coincides with
the one given for the non-abelian branes, see
for example  Eq. (\ref{bou73ba}), with the
identification $q_{(1)}=N_{1}$. The only difference
between the two is in the phase factor written in Eq. (\ref{D25ph}).
As already stressed in the Introduction, this factor
does not influence the one-loop free energy.

We can consider the more general case in which the T-duality
acts on each torus $T^2_{(\alpha)}$ as in Eq. (\ref{Lambda2}) with
parameters $p_{(\alpha)}, q_{(\alpha)}$, getting:
\begin{eqnarray*}
|D25(E,F)\rangle_{z m}\!\! &=&\!\!\prod_{\alpha=1}^{\frac{\hat
d}{2}}\left[q_{(\alpha)} \frac{\sqrt{\det {\cal E}_{(\alpha)}}
}{\left(\det G_{(\alpha)}\right)^{1/4}}\sum_{s^{2\alpha},\,s^{2\alpha-1}\in Z} | n_{2\alpha-1}= p_{(\alpha)} s^{2\alpha},n_{2\alpha}=- p_{(\alpha)} s^{2\alpha-1}\rangle \right.\nonumber\\
&\times& \!\!\!\left. |m^{2\alpha-1}= q_{(\alpha)} s^{2\alpha-1},\,
m^{2\alpha}= q_{(\alpha)} s^{2\alpha}\rangle\right].
\label{D25zm-fact}
\end{eqnarray*}

{\em Plain $Dp$ branes}
\label{sect-plain-Dp}

In order to make contact with the kind of T-duality that transforms
Neumann into Dirichlet boundary conditions and viceversa, we consider
another particular case of the Narain T-duality group. We still consider
a non-magnetized space filling  brane, and on it we act  with
the  special  case of the standard T-duality, given by:
\begin{equation}
\Lambda=\left(\begin{array}{cccc}
\uno_p &    &   0_p    &  \\
    &     0_{d-p}      &      & \uno_{d-p}   \\
     0_p &  & \uno_p &  \\
      & \uno_{d-p} & & 0_{d-p}\\
\end{array}
\right) \label{usut}
\end{equation}
with $d=25$. This T-duality transforms a plain $D25$ into a
configuration of a plain $D p$
\begin{eqnarray*}
&&{ P}_i^t |D25(E^t,F^t=0)\rangle = 0
\nonumber\\
\Rightarrow && P_{i_\parallel} |Dp (E,F=0)\rangle = X'^{i_\perp} |Dp
(E,F=0)\rangle =0
\end{eqnarray*}
with $i_\parallel= 1,\dots p$ and $i_\perp=p+1,\dots, d$.
 Let us give the transformation property of the boundary state under
 the T-duality transformation in Eq. (\ref{usut}). The non-zero
modes of the boundary state transform according to Eq.
(\ref{B-nzm-can}) with
\begin{equation*}
(\C + \D E)^{T} (-\C+\D E^{ T})^{-T} = \left(\begin{array}{c c}
(E_{\parallel~\parallel})^T(E_{\parallel~\parallel})^{-1}  &
0 \\
2 E_{\perp~\parallel}(E_{\parallel~\parallel})^{-1}  & -\uno_{\perp
\perp}
\end{array} \right)
\end{equation*}
where $E_{\parallel~\parallel}= \parallel E_{i_\parallel
  j_\parallel}\parallel$,
$E_{\parallel~\perp}= \parallel E_{i_\parallel  j_\perp}\parallel$,
and so on. The zero modes transform according to Eq.
(\ref{inverse-mn-a}). Thus the boundary state becomes:
\begin{eqnarray*}
| Dp(E,F=0) \rangle &=& \frac{T_{25}}{2}
 \frac{\sqrt{\det_p { E_{\parallel~\parallel}}}}{\left(\det G\right)^{1/4}}~
\sum_{s\in Z^{25}} | n_\parallel=0, n_\perp=s_\perp,
m^\parallel=s^\parallel, m^\perp=0 \rangle
\nonumber\\
&\times& e^{-\sum_{n=1}^\infty \frac{1}{n}
~\alpha^{i_\parallel}_{-n}(E^T_{i_\parallel k_\parallel}
 (E^{-1})^{k_\parallel h_\parallel}G_{h_\parallel j_\parallel}
\alphat^{j_\parallel}_{-n} }
\nonumber\\
&\times& e^{-2 \sum_{n=1}^\infty \frac{1}{n} ~\alpha^{i_\perp}_{-n}
{E^T}_{i_\perp k_\parallel}(E^{-1})^{k_\parallel h_\parallel} (
G_{h_\parallel j_\parallel}\alphat^{j_\parallel}_{-n} }
\nonumber\\
&\times& e^{+\sum_{n=1}^\infty \frac{1}{n} ~\alpha^{i_\perp}_{-n}
G_{i_\perp j_\perp} \alphat^{j_\perp}_{-n} } ~| 0_a, 0_{\tilde a}
\rangle
\nonumber\\
&\times& e^{-\sum_{n=1}^\infty \frac{1}{n} ~\alpha^{0}_{-n} G_{0 0}
\alphat^{0}_{-n} } ~| k_0=0 \rangle \label{Dpzmt}
\end{eqnarray*}
where the terms with
$E_{\perp~\parallel}^T(E_{\parallel~\parallel})^{-1} $ are present
since the reflection conditions\\ $P_{\parallel}|Dp
(E,F=0)\rangle=0$ mix both $\alpha^{\parallel}$ and
$\alpha^{\perp}$.

\subsection{General transformation of $F$ under T-duality}
\label{FunderT}

We want now to determine how $F$ and $\Theta$ transform under
T-duality.
Let us start from a $D25$ brane with a constant $F$
\begin{equation*}
\left[{P}_i -  \hat F_{i j} \frac{X'^{ j}}{2\pi\alpha'}\right]|D25(E,F)\rangle
= 0
\end{equation*}
and then perform a T-duality transformation given by the matrix
$\Lambda^{-1}$ in
Eq. (\ref{Lambda-1}). In so doing we get:
\begin{eqnarray*}
\left[(\A^T - \hat F \B^T)_{i j}{P}^t_j +   (\C^T - \hat F \D^T)_{i
j}\frac{X^{t~'~ j}}{2\pi\alpha'}\right]|D25(E^t,F^t)\rangle = 0
\label{trainv}
\end{eqnarray*}
 It is then easy to find that  when
\begin{equation*}
 \det ( \A + \B \hat F)\ne 0
\end{equation*}
the system still describes a $D25$ brane
\begin{eqnarray*}
&&[{P}^t_i -  \hat F^t_{i j} \frac{X^{t~'~
j}}{2\pi\alpha'}]|D25(E^t,F^t)\rangle = 0 . \label{bc2b}
\end{eqnarray*}
Indeed if ${\rm det}\left(\A+{\hat F}\B \right) \neq 0$, Eq.
(\ref{trainv}) can be written as:
\begin{eqnarray*}
&&\left( \A+  \B  {\hat{F}} \right)^T \left[ P^t +\left( \A+  \B
{\hat{ F}} \right)^{-T} \left(\C+ \D {\hat{F}} \right)^T
\frac{{X'}^t}{2\pi\alpha'}\right]|D25(E^t,F^t)\rangle = 0.
\end{eqnarray*}
By comparing this equation with Eq. (\ref{bc2b}), we get:
\begin{eqnarray}
&&\hat F^t =  - (\A^T - \hat F \B^T)^{-1} (\C^T - \hat F \D^T)=( \C
+ \D \hat F) ( \A + \B \hat F)^{-1}.  \label{Ft-F}
\end{eqnarray}
Notice that
\begin{eqnarray}
{\hat F}^t=0\Rightarrow
{\hat F}={\hat D}^{-1}{\hat C}={\hat C}^T{\hat D}^{-T}.
\label{FtF}
\end{eqnarray}
On the other side,  when $\det ( \A + \B \hat F)=0$, some directions
acquire Dirichlet boundary conditions.  In particular when
\begin{equation*}
\A + \B \hat F=0
\end{equation*}
Eq. (\ref{trainv}) reduces to
\begin{eqnarray*}
&& X^{t ' i}|D1(E^t)\rangle = 0
\end{eqnarray*}
corresponding to pure Dirichlet boundary conditions.

When $
 \det ( \A + \B \hat F)\ne 0
$ we can then evaluate how $\cal E$ transforms:
\begin{eqnarray*}
{\cal E}^t &=& E^{t T}+ \hat  F^t =(\A +\B E)^{-T}(\C+\D E)^T+ ( \C
+ \D \hat F) ( \A + \B \hat F)^{-1}
\nonumber\\
\end{eqnarray*}
where we have used Eq.s (\ref{Et-E}) and (\ref{Ft-F}), together with
Eq.s (\ref{LLT}) and (\ref{LTL}). One can equivalently write:
\begin{eqnarray}
{\cal E}^t
&=&
=E^{t T}+ \hat{F}^t = (\A + \B \hat F)^{-T} {\cal E}(\A - \B E^T)^{-1}.
\label{calEt-calE}
\end{eqnarray}
{From} them we deduce
\begin{eqnarray}
{\cal G}^{t~-1} &=& (\A  + \B \hat F ) {\cal G}^{-1} (\A  + \B \hat
F )^T = T(F)~ {\cal G}^{-1} ~T^T(F) \label{calGt-calG}
\\
\Theta^t &=& (\A  + \B \hat F ) \Theta (\A  + \B \hat F )^T + \B (\A
+ \B \hat F )^T = T(F)~ \Theta ~T^T(F) + \B T^T(F)
\nonumber\\
\label{Thetat-Theta}
\end{eqnarray}
with  ${\cal G}$ and $\Theta$ defined in appendix
\ref{app-conventions}. Here we have introduced the matrix
\begin{equation}
T(F)= (\A  + \B \hat F ) = ( \D - F^t \B )^{-T} =T^{t~-1}(F^t)
\label{Tt-T}
\end{equation}
and used the relation:
\begin{equation*}
T(F)\B^T+\B T^T(F)=0. \label{TB}
\end{equation*}
Notice the important fact that $\Theta$ does {\bf not} transform
``tensorially'' under a T-duality but it behaves like a connection
and has a shift term.

The inverses of the previous equations can be obtained using Eq.
(\ref{Lambda-1}), i.e $\A\leftrightarrow \D^T, \B\rightarrow \B^T,
\C \rightarrow \C^T$ and exchanging $^t$ quantities with those
without $^t$.
In comparing this set of equations with the one we have written we
find
\begin{equation*}
( \A - \B E^T) = (\D -E^t \B)^{-T} ,~~~~ ( \A + \B E ) = (\D -E^{t
T} \B)^{-T} \label{A-BET}
\end{equation*}

\subsection{Adding Wilson lines to the boundary state}

In section \ref{F=0} we have seen that a  $D$-brane with $F=0$ wrapped $N$-times  around a torus
 can be described as a brane with gauge group $U(N)$ and a Wilson line background which induces a non trivial holonomy \cite{Hashimoto:1996pd}.
 The boundary state description for such a brane can therefore be obtained
turning on Wilson lines to the  boundary state discussed in section
\ref{nabrplbr} in the special case $F^t=0$. More explicitly  we
have to compute:
\begin{eqnarray*}
|D25(E^t,F^t=0,a^t )\rangle &=& {\rm Tr}~ P e^{-i q \int_0^\pi
d\sigma~ a_i^{t} X'^{t~i} } |D25(E^t,F^t=0)\rangle
\nonumber\\
&=& \sum_{b=1}^N
 e^{-i 2\pi \sqrt{\alpha'} q a^{t~b}_{ i} \hat
 m^i}|D25(E^t,F^t=0)\rangle\nonumber\\
 &=& \frac{T_{25}}{2} \frac{\sqrt{\det { { E}^t}}}{\left(\det
G^t\right)^{1/4}}~
 |D25(E^t,F^t=0)\rangle _{n z m}
\nonumber\\
&&\times \sum_{b=1}^N \sum_{s\in Z^{25}} e^{-i 2\pi \sqrt{\alpha'}
q a^{t~b}_i s^i } | n^t_i= 0, m^{t~i}= s^i\rangle. \label{D25t+wilson}
\end{eqnarray*}
Let us also explore the effect of a T-duality transformation on the Wilson line itself.
By performing a T-duality on the previous expression one is
immediately led to:
\begin{eqnarray*}
|D25(E,F,a)\rangle &=& \frac{T_{25}}{2} \sqrt{\det \D}
\frac{\sqrt{\det { \cal E}}}{\left(\det G\right)^{1/4}}~
 |D25(E,F)\rangle _{n z m}
\nonumber\\
&&\times \sum_{b=1}^N \sum_{s\in Z^{25}} e^{-i 2\pi \sqrt{\alpha'}
q a^{t~ b}_i s^i } | n_i= {\C^T}{}^i_{~j}s^j, m^{i}=
{\D^T}{}^i_{~j}s^j\rangle \label{D25t+wilson-Tdual}.
\end{eqnarray*}
Since we expect the Wilson line to be multiplied by the winding $
m^{i}= {\D^T}{}^i_{~j}s^j$ we deduce that
\begin{equation}
a^{t~b}_i  {\D^{-T}}{}^i_{~j} = a^{b}_j \Rightarrow a^{t~ b} T(F) =
a^{b}
\end{equation}
where we have used Eq. (\ref{Tt-T}) for the special case $F^t=0$ and
assumed that the transformation must be dependent on $F$.

In order to extend the previous discussion to the case in which both a
Wilson line background  and an abelian background gauge field $F$ (proportional to the unity) are turned on,
one can follow the same procedure since
$A_i(x)+a_i$ can be added to the boundary either as $tr~ P e^{-i q
\int_0^\pi d\sigma~ (A_i+a_i) X'^{t~i} } $ or in two steps by
computing $ tr~ e^{-i q \int_0^\pi d\sigma~ a_i X'^{t~i} } ~e^{-i
q \int_0^\pi d\sigma~ A_i  X'^{t~i} } $. By choosing  the second
procedure,
one starts with a boundary with $F^t=0$, then one constructs
the one with $F\neq 0$ by means of  a T-duality
transformation with matrix $\Lambda_0$  and finally one  adds Wilson
lines.
In this way one gets:
\begin{eqnarray}
|D25(E,F,a)\rangle &=& tr~ P e^{-i q \int_0^\pi d\sigma~ a_i X'^i}
|D25(E,F)\rangle
\nonumber\\
&=& \sum_{b=1}^N
 e^{-i 2\pi \sqrt{\alpha'} q a^{b}_{ i} \hat m^i}|D25(E,F)\rangle
\nonumber\\
&=& \frac{T_{25}}{2} \sqrt{\det \D_0} \frac{\sqrt{\det { \cal
E}_0}}{\left(\det G\right)^{1/4}}~
 |D25(E,F)\rangle _{n z m}
\nonumber\\
&&\!\!\times \sum_{b=1}^N \sum_{s\in Z^{25}} e^{-i 2\pi \sqrt{\alpha'}
qa^{b}_i {\D_0^T}{}^i_{~j}s^j } | n_i= {\C^T_0}{}^i_{~j}s^j, m^{i}=
{\D^T_0}{}^i_{~j}s^j\rangle \label{D25+wilson}
\end{eqnarray}
where $\hat F =\C_0^T \D_0^{-T}$, $ |D25(E,F)\rangle _{n z m}$ is
the non zero mode part of the boundary, and $ \hat m^i$ is the
winding operator.

Finally one can study the transformation properties of the Wilson lines under T-duality   in the case
$F\neq 0$. Performing a second T-duality transformation with matrix
$\Lambda$ (see (\ref{CanTras})) on the previous boundary, its zero
mode part becomes  (see Eq. (\ref{mtnt-mn}))
\begin{eqnarray*}
|D25(E^t,F^t,a^t )\rangle_{z m}&=&\sum_{b=1}^N
 \sum_{s\in Z^{25}}
e^{-i 2\pi \sqrt{\alpha'} q a^{b}_{ i} {\D_0^T}{}^i_{~j}s^j }\nonumber\\
&\times&| n^t=
(\C \D^T_0+\D \C^T_0) s, m^t= (\A\D^T_0+\B\C^T_0) s\rangle
\nonumber\\
\end{eqnarray*}
from which we get the transformation rule of the Wilson line under
T-duality:
\begin{equation*}
a^{T}_b \D_0^T = a_{b}^{t T} (\A\D^T_0+\B\C^T_0) \Rightarrow a_{b}^T
= a_{b}^{t T} T(F) \label{at-a}
\end{equation*}
which is valid for the case $\det T(F)\ne 0$; when $\det T(F)= 0$
there are directions $x^d$ with DD boundary conditions where it is not
possible to add Wilson lines anymore as in (\ref{D25+wilson}) but it is
possible to move the brane by $e^{-i  \int_0^\pi d\sigma~ \Delta_d
\dot X^d}$.

\subsection{Vertex operators and scattering amplitudes in Narain branes}
\label{osv}

In the previous subsections we have studied how the boundary state and
a gauge field ${\hat{F}}$ living on a D25  brane
transform under the most general
T-duality transformation. In particular we have seen that, starting
from a configuration without gauge fields on the
branes, the T-duality transformation, performed by the matrix
$\Lambda$ given in Eq. (\ref{lam}), provides a configuration with
a non-zero gauge field
given  in Eq. (\ref{FtF}).  Moreover also the
 boundary state acquires the same gauge field (Eq. (\ref{D25})). In
 this way one can obtain a theory with a non-zero gauge field from
a theory without it. Of course, if we transform not only the operators but
 all quantities appearing in a string amplitude, as for instance the momenta of the external
particles, nothing will change
because T-duality is a symmetry of string theory.

In this section we want to extend this procedure to the vertex operators.
Since we know that we are  going to get a theory with a gauge field given by
$\hat F= \C^T \D^{-T}$,  we can immediately write
the vertex operators describing the emission of open string
tachyons. They are given by:
\begin{eqnarray}
{\cal V}_{(0)}(x; k)
&=&  e^{iD_0(k,\hat F, B;\hat p)}
:e^{ i \left(k_0 X^0_{L}(x)+k_iX^i_{L (0)}(x)\right)} :
\nonumber\\
{\cal V}_{(\pi)}(y; k)
&=& e^{iD_\pi(k,\hat F, B;\hat p)}
: e^{ i \left( k_0 X^0_{L}(x)+k_i X^i_{L (0)}(y)\right)} :
\label{verticiap}
\end{eqnarray}
where we assume that all the spatial directions are
compactified. Here  $x=|x|$, $y=|y| e^{i \pi}$.
We consider only one of such branes and consequently we do not need
to introduce any Chan-Paton factor.
The factors $e^{iD_{0,\pi}(k,\hat F,B;\hat p)}$ are the cocycles phase
factors\cite{Pesando:2003} and are necessary to have a
well-defined theory of open and closed strings. They can be explicitly
derived by requiring the theory to satisfy certain
specific constraints such as the commutativity among open
and  closed string vertices and
a proper behavior of the vertices under Hermitian conjugation.
Cocycles play an important rule in determining whether
the theory is commutative or not. However, their explicit knowledge is
not crucial for discussing  the main features of Narain
branes and for comparing them  with the
non-abelian bundle description of magnetized branes.
Therefore we postpone their explicit evaluation and the discussion about the
commutation property of the theory to further developments.

Before going on, let us now spend a few words about how
Eq. (\ref{verticiap}) can be derived. The vertex operator for an
open string tachyon is obtained by inserting
in the exponent of the vertex operator the
string coordinate in Eq. (\ref{icso}), computed at one of the endpoints
of the open string. But then, for instance in the case of $\sigma =0$,
one can use Eq.s (\ref{icsot}) and
express it in terms of $X^{i}_{(0)L}$, apart from
a phase that contributes to the cocycle that we
are anyway not considering as
explained above. A similar reasoning can also be used
for the other endpoint.

Let us  now consider the vertex for the closed  string tachyon. If the vertex
has to be used in an amplitude with only closed strings
(sphere diagram), then, apart from a
cocycle factor,  it is given by:
\begin{eqnarray}
{\cal{W}}_{T_c} ( z , {\bar{z}}; k_L , k_R ) = : e^{(i k_{0}   X^{0} (z, {\bar{z}} ) +
k_{Li}  X^{i}_{L} (z)  + k_{Ri}  {\tilde{X}}^{i}_{R} ({\bar{z}}) ) }:
\label{closeve}
\end{eqnarray}
where $X_L$ and ${\tilde{X}}_R$ are given respectively in Eq.s (\ref{xi75}) and (\ref{xi76}) and the variables $z$ and ${\bar{z}}$ are defined in the entire complex plane.

On the other hand,  the vertex for the closed string tachyon
describing  interactions on the disk diagram is given
by~\cite{Ademollo,9702037,Pesando:2003}
\begin{eqnarray}
&&{\cal W}_{T_c }(z,\bar z; k_L,k_R,y_0)
=e^{iD_C(k,\hat F, B;\hat p)}\nonumber\\
&&:e^{ i \left[\frac{1}{2} k_0 X^0_{L}(z) +
k_{L i}(G^{-1} {\cal E})^i_{\,j} X^j_{L (0)}(z)\right]} :
:e^{ i \left[\frac{1}{2}k_0 X^0_{R}(\bar z)+
k_{R i}(G^{-1} {\calE}^T)^i_{\,j} X_{R (0)}^i(\bar z) \right]} :
\label{main-OpenStringVertices}
\end{eqnarray}
Here $z$ is defined in the upper half  complex plane
and the phase factor $e^{iD_C(k,\hat F, B;\hat p)}$
are the closed string cocycles. After having determined the open and
closed string vertex operators, one can then compute the scattering amplitudes
involving them, but, before doing that, let us first describe the action of
T-duality on both the closed and open string vertex
operators. This will allow us to rederive Eq.s (\ref{verticiap})
and, more importantly, to study in which sense the Narain branes
are wrapped branes.

Let us start from the closed string vertex given in Eq. (\ref{closeve})
and show that, under a T-duality transformation, it keeps
the same form. By considering of course only the compact space part
and extending the T-duality transformations in Eq.s (\ref{hol-antihol-trans})
to be valid also for
the coordinates $X_{L}^{i}$ and ${\tilde{X}}_{R}^{i}$, and not just
for their derivatives, we get:
\begin{eqnarray*}
X_{L}^{t i} = ( {\cal{A}}+ {\cal{B}} E ) X_{L}^{i}
~~;~~
X_{L}^{t i} = ( {\cal{A}}- {\cal{B}} E^T ) {\tilde{X}}_{R}^{i}.
\label{TduaX}
\end{eqnarray*}
On the other hand, we have to remember that also the external momenta
$k_L$ and $k_R$ transform according to Eq.s (\ref{p-pr-transf}) that,
with Eq.s  $k_{L,R\,i}=G_{ij}k^j_{L,R}$ and (\ref{Gt-G}),
imply the following  transformations:
\begin{eqnarray}
&&
k^{~T}_{Li} = k^{t~T}_{Lj} (\A + \B E)^{j}_{~i}
~~~~
k^{~T}_{Ri} = k^{t~T}_{Rj} (\A - \B E^T)^{j}_{~i} .
\label{cappa}
\end{eqnarray}
By using the two previous equations, it is easy to see that the exponent in the
vertex operator remains the same in form:
\begin{eqnarray*}
k^{t~T}_{Li} X_{L}^{t i}  + k^{t~T}_{Ri} {\tilde{X}}_{R}^{t i}  =
k^{T}_{Li} X_{L}^{ i}  + k^{T}_{Ri} {\tilde{X}}_{R}^{ i}
\label{steform}
\end{eqnarray*}
where the index $T$ has been introduced here for the sake of clarity.

Let us consider now the vertex for the closed string tachyon to be used
on a disk amplitude given in Eq. (\ref{main-OpenStringVertices}).
{From} the transformation properties of the closed string momenta
given in Eq.s (\ref{cappa}) one can
deduce the transformations under T-duality of the  left and
right moving parts of the vertex operator by
requiring that the closed string vertices, written in open string
formalism, are invariant in form under such a transformation.
In this way we get:
\begin{eqnarray}
G^{t~-1} {\cal E}^t X^t_{L (0)}(z)&=&(\A + \B E) G^{-1} {\cal E} X_{L (0)}(z),
\nonumber\\
G^{t~-1} {\cal E}^{t~T}  X^t_{R (0)}(\bar z)
&=& (\A - \B E^T) G^{-1} {\cal E}^T X_{R (0)}(\bar z)
\label{trax}
\end{eqnarray}
for the left and right components of $X$.

But we have to take into account that in the open string formalism
the left and right parts are not independent because
of the reflection conditions. They can be of two types either
Neumann or Dirichlet.

One can therefore distinguish two cases:
\begin{enumerate}
\item
The reflection conditions are generalized Neumann boundary conditions\footnote{
We call these b.c. generalized Neumann because they are Neumann
b.c. on the $X_{(0)}$ fields but not on $X$ ones; this in the spirit
of the asymmetric rotation of \cite{Blumenhagen:2000fp}.
} in both the original
theory and in the T-dual one. This implies  the two equations:
\begin{equation}
X^t_{L (0)}(x)= X^t_{R (0)}(x)
~~;~~
X_{L (0)}(x)= X_{R (0)}(x) .
\label{boundcond}
\end{equation}
Thus, after imposing $X^t_{L (0)}(x)= X^t_{R (0)}(x)$
and using $X_{L (0)}(x)= X_{R (0)}(x)$ in Eq. (\ref{trax}),  one gets:
\begin{eqnarray*}
\left( G^{t~-1} {\cal E}^t \right)^{-1} (\A+\B E) G^{-1} {\cal E}
&=& \left( G^{t~-1} {\cal E}^{t~T} \right)^{-1} (\A-\B E^T) G^{-1}
{\cal  E}^T\nonumber\\
&=& \A + \B F
= T(F).
\end{eqnarray*}
The third identity can be verified by using Eq.s (\ref{Gt-G}) and (\ref{calEt-calE}).
{From}  Eq.s (\ref{trax}) we find:
\begin{equation}
X^t_{L (0)}(z)= T(F)~ X_{L (0)}(z)
~~~~
X^t_{R (0)}(\bar z)= T(F)~ X_{R (0)}(\bar z)
\label{Xt-X-y0t-y0}
\end{equation}
which can be simply interpreted as due to the fact that in the T-dual system
distances are rescaled by $T(F)$.
Thus the boundary conditions in Eq. (\ref{boundcond}) can both be
imposed when
\begin{equation*}
\det T(F) = \det ( \A + \B \hat F) \ne 0 .
\end{equation*}
After having  examined the closed string vertices, we
can now  discuss  what happens to the open string
vertices in Eq.s (\ref{verticiap}). If we require them to remain
invariant in form, the following equation has to be imposed:
\begin{eqnarray*}
  e^{ i k^{t~T} X^t_{L (0)}(x)  }
=  e^{ i k^{t~T}  T(F)~ X_{L( 0)}(x) } = e^{ i k^{T} X_{L (0)}(x)  }
\label{opeinv}
\end{eqnarray*}
where we have used Eq.  (\ref{Xt-X-y0t-y0}) and
\begin{eqnarray}
k^T =  k^{t~T}  T(F).
\label{calMt-calM}
\end{eqnarray}
This is the vertex that we have already written down in
Eq. (\ref{verticiap}).
The transformations in Eq. (\ref{Xt-X-y0t-y0}) are obviously
consistent with the
OPEs in  Eq. (\ref{openOPEs-X0}) when
\begin{equation*}
\calG^t=T^{-T}(F) \calG T^{-1}(F)
\end{equation*}
which matches perfectly the first equation  in (\ref{calGt-calG}).

\item
 The reflection conditions are generalized Neumann boundary conditions in the original
  theory  and mixed generalized Neumann and Dirichlet boundary conditions in the T-dual
  one. This is the case when
\begin{equation*}
\det T(F) = \det ( \A + \B \hat F)= 0 .
\end{equation*}
This condition is very general and corresponds to various different
generalized Neumann and Dirichlet boundary conditions according to the number of
zero eigenvalues of the matrix $T$. For our discussion we consider a
special subcase characterized by the following condition:
\begin{equation*}
T= \A + \B \hat F= 0  \Rightarrow \hat F^t=\infty.
\end{equation*}
This special case corresponds to the Dirichlet reflection conditions
in all the compact $X^t$ coordinates:
\begin{equation}
X^t_{L (0)}(x)= -X^t_{R (0)}(x)
~~;~~
X_{L (0)}(x )= X_{R (0)}(x ).
\label{bounda2}
\end{equation}
However in this case the T-dual coordinate does not have
anymore the expansion in Eq.s
(\ref{icsol}) and (\ref{icsor}), but one simply has (up to cocycles)
\begin{eqnarray*}
{\hat X}^t_L(z)= X^t_{L (0)}(z)~~;~~
{\hat X}^t_R(\bar z)= X^t_{R (0)}(\bar z).
\label{conty}
\end{eqnarray*}
Comparing the vertices in the two T-dual theories as in
Eq. (\ref{calMt-calM}) yields:
\begin{eqnarray*}
 X^t_{L(0)}(x)&=&(\A+\B E) G^{-1} \calE X_{L(0)}(x);\nonumber\\
 X^t_{R(0)}(x)&=&(\A-\B E^T) G^{-1} \calE^T X_{R(0)}(x).
\end{eqnarray*}
Hence, the boundary conditions in Eq. (\ref{bounda2}) are consistent if the following equation holds:
\begin{equation*}
 (\A+\B E) G^{-1} \calE = - (\A-\B E^T) G^{-1} \calE^T= \B \calG.
\end{equation*}
This can be verified with the help of  $\hat F= -\B^{-1} \A$.
Therefore we get a relation between the
``momentum'' (actually distance) in Dirichlet directions $k^t$ and the
momentum $k$, given by:
\begin{equation*}
k^T = k^{t T} \B \calG.
\end{equation*}

\end{enumerate}
Moreover, in both cases, by using
Eq.s (\ref{hol-antihol-trans}) and (\ref{Gt-G}), we have
\begin{equation*}
T(z)
=- \frac{1}{\alpha'}  \partial X_L^T(z)  G  \partial X_L(z)
=- \frac{1}{\alpha'}  \partial X_L^{t T}(z)  G^t  \partial X_L^{t}(z)
\end{equation*}
and, because of this, all conformal dimensions are preserved.

We are now ready to compute the scattering amplitudes involving open
and closed strings in the case of non-abelian and Narain branes, and
compare them. As we will see below, the amplitudes exhibiting a
difference between the non-abelian and Narain branes are the ones
involving both open and closed strings. In particular,  it is
sufficient to limit ourselves to external tachyons. We only consider
the compact part of correlators involving one closed  string tachyon
and  $M$ open string tachyons, up to phases that we have
systematically neglected in this paper. In the case of non abelian
branes one gets

\begin{eqnarray}
&&
\langle 0|
 W(z, \bar z; k_{L }, k_{R })
V(x_1; k_1)
\dots
V(x_M; k_M)
|0 \rangle_{compact}
\nonumber\\
&=&
\mbox{Tr} \left[ \prod_{i=1}^M \Lambda_{(k_1^{(i)}, k_2^{(i)})} \right]
A(z, \bar{z}, x_{r}, k_{i})
\delta_{\calE^{T} G^{-1} k_L+\calE G^{-1} k_R +\sum_{r=1}^{M} k_r,0}
\label{Nonab}
\end{eqnarray}
 while for the Narain branes one gets
\begin{eqnarray*}
&&
\langle 0|
 W(z, \bar z; k_{L }, k_{R })
V(x_1; k_1)
\dots
V(x_M; k_M)
|0 \rangle_{compact}
\nonumber\\
&=&
A(z, \bar{z}, x_{r}, k_{i})
\delta_{\calE^{T} G^{-1} k_L+\calE G^{-1} k_R +\sum_{r=1}^{M} k_r,0}
\label{1-Tc-M-To}
\end{eqnarray*}
where
\begin{eqnarray}
A(z, \bar{z}, x_{r}, k_{i}) &\equiv&
\prod_{r=1}^M
(z-x_r)^{2\alpha' k_L^T G^{-1} \calE \calG^{-1} k_r}
(\bar z-x_r)^{2\alpha' k_R^T G^{-1} {\calE}^T \calG^{-1} k_r}
\nonumber\\
& \times &
\prod_{1=r < s}^M (x_{r}-x_{s})^{2\alpha' k_{r}^T \calG^{-1} k_{s}}
\left(z -\bar z\right)^{2\alpha' k_L^T \calE^{-T} \calE G^{-1} k_R }.
\label{1-Tc-M-To2}
\end{eqnarray}
Eqs. (\ref{Nonab}) and (\ref{1-Tc-M-To}) are easily seen to differ only for
the trace over the Chan Paton factors which is given in Eq. (\ref{trace2}).
This is the  consequence of the fact that
the vertex operators for closed string tachyons is the one given
in Eq. (\ref{closeve}) for both theories, while the ones
for open string tachyons have
the same operatorial part as those in Eq. (\ref{verticiap}), but
those  in the non-abelian branes  have in addition momentum
dependent Chan-Paton factors, given in Eq. (\ref{cpfndc}).

By using the formulas in the Appendices it is easy to show that
the $\delta$-function which is common to the two correlators gives:
\begin{eqnarray*}
n_i - {\hat{F}}_{ij}  m^j + \sum_{r=1}^{M} \frac{n_{i}^{r}}{N} =0~~;~~i=1,2
\label{deltaco}
\end{eqnarray*}
in terms of the momentum $n_i$ and winding number $m^i$ of the
closed string and the momenta
$ \frac{n_{i}^{r}}{N} $ of the open strings. For the sake of simplicity we
have
restricted our analysis to the two direction of a  torus $T^2$.
Using Eq. (\ref{chernc89}) in  the previous equation one gets:
\begin{eqnarray}
n_1 - \frac{m m^2}{N} + \sum_{r=1}^{M} \frac{n_{1}^{r}}{N} =
n_2 + \frac{m m^1}{N} + \sum_{r=1}^{M} \frac{n_{2}^{r}}{N}=0
\label{n1n2}
\end{eqnarray}
They can be satisfied only if the following relations holds
\begin{eqnarray}
\sum_{r=1}^{M}  n_{1}^{r} -m m^2 = s_1 N~~;~~
\sum_{r=1}^{M}  n_{2}^{r} +m m^1 = s_2 N
\label{s1s2}
\end{eqnarray}
where $s_1$ and $s_2$ are arbitrary integers. Finally inserting them
back in Eq. (\ref{n1n2}) one gets:
\begin{eqnarray}
n_1 + s_1 = n_2 + s_2 =0
\label{n1s1}
\end{eqnarray}
The  constraint imposed by the $\delta$-function
can be satisfied only if Eq.s (\ref{s1s2})
and (\ref{n1s1}) are satisfied

In the case of non-abelian branes the trace over the Chan Paton factors imposes the following additional constraints:
\begin{eqnarray*}
\sqrt{\alpha'} N \sum_{r=1}^{M} k_{1,2}^{r} = \sum_{r=1}^{M} n_{1,2}^{r} =
 r_{1,2} N
\label{conex}
\end{eqnarray*}
where $r_{1,2}$ are integer numbers.

In conclusion, for non-abelian branes the relations in
Eq. (\ref{conex}) must be considered together with Eq.s  (\ref{s1s2})
and (\ref{n1s1}).  Therefore the class of solutions that one gets
for the Narain branes is bigger than the one for the non-abelian branes
and this means that
the two theories are not equivalent. Notice, however, that a difference can
be noticed  only if the scattering amplitude
involves at least one closed string. If we had only open strings then
one would get precisely the same conditions for the two cases. The same
is true for the case of a closed string with $n_i =m^i =0$.

In the previous section we have shown that the non-abelian branes provide
a description of branes wrapped $N$ times on the torus $T^2$ through the
introduction of a non-abelian gauge bundle based on
the gauge group $U(N)$. In other words, the wrapping number
$N$ is provided by the order of the gauge group. This is the reason
why for this kind of branes we must introduce Chan-Paton factors that turned
out to be momentum dependent.
In the case of the Narain brane we do not have any non-abelian gauge field.
Then in what sense do the Narain branes provide a description of
wrapped branes? Or
can we say that the Narain branes provide an alternative description
of them? And if yes, what is the precise meaning to give to this claim?

In the introduction we have discussed two possibilities for
obtaining Eq. (\ref{chernc89}). The first one is based on the
presence of a non-abelian gauge bundle and this is the one realized
by the non-abelian branes. In the following, we aim to show that the
Narain branes seem to realize the other possibility discussed around
Eq. (\ref{chernc89a}). In order to see how this comes about, we have
to study what happens to the open string coordinate when we go
around the torus. This is what we are going to discuss in the last
part of this subsection.

We start  rewriting the tachyon open string vertex operator in
Eq. (\ref{verticiap}) for a Narain
brane which is obtained
through a T-duality from a plain brane with $F^t=0$, as follows:
\begin{equation*}
V_{(0) T, c}(x;k) \sim :e^{i k^t_i (\D^{-T})^i_{~j} X^{j}_{L (0)}(x)} :
\label{generic-Narain-vertex}
\end{equation*}
where we used $T(F)=\D^{-T}$ (because of Eq. (\ref{Tt-T})).
The fact that $\sqrt{\alpha'} k^t_i\in \Z$
immediately implies that the theory is
invariant under
\begin{equation*}
X  \rightarrow X + 2\pi \sqrt{\alpha'} \D^T s
~~~\forall s\in Z^{25}
\label{NarainBranePeriod}
\end{equation*}
while in the original theory the vertex operator was only invariant under
\begin{eqnarray*}
X^t  \rightarrow X^t + 2\pi \sqrt{\alpha'}  s.
\label{Xt}
\end{eqnarray*}
The periodicity of the open string coordinate
can also be verified directly starting from the
operator which performs a shift of $X^t\rightarrow X^t+2\pi s$ and
rewriting it in the T-dual theory
\begin{equation*}
{\cal T}_s^t
= e^{2\pi i ~s^T {\cal G}^{t } p^t}
= e^{2\pi i ~s^T  {\cal G}^t T(F)  p}
= e^{2\pi i ~s^T  T^{-T}(F) {\cal G}   p}
= e^{2\pi i ~(\D^T s)^T  {\cal G}  p}
=
{\cal T}_{\D^T s}.
\end{equation*}
In order to see more explicitly what happens it is convenient
to  specialize the previous discussion to the first case
treated in Section \ref{T-dual-fact-torus-B}.
There, T-duality acts on each torus $T^2_{(t)}$ as in
Eq. (\ref{Lambda2}) with parameters $(p_{(t)}, q_{(t)})$.
It is straightforward to write the compact part of the tachyonic vertex, in fact, by focusing only on the first torus:
\begin{equation*}
V_{(0) T, c}(x;k) \sim
:e^{i \frac{ n_{1}  X^{1}_{L (0)}(x)+n_2  X^{2}_{L (0)}(x)}{\sqrt{\alpha'}~q_{(1)}}}:
\label{Narain-fact-D25vertex}
\end{equation*}
where the compact momentum is
$\sqrt{\alpha'}~k=\left(\frac{n_{1}}{q_{(1)}},\frac{n_{2}}{q_{(1)}}\right)$
with all $n$ integers.

Unlike the open string vertices in the case of non-abelian branes, these
vertices have no  Chan-Paton factors  and  describe
objects with non-trivial wrapping
\begin{eqnarray*}
X^{1,2}= X^{1,2}+2\pi\sqrt{\alpha'}q_{(1)}s.
\end{eqnarray*}

As discussed in section \ref{T-dual-fact-torus-B}, the normalization factor in front of the boundary state in Eq. (\ref{D25zm-fact-1}) suggests that a Narain brane is a brane wrapped $q$ times around the whole torus. This means that the
previous periodicity conditions have to be interpreted as simultaneous conditions on $X^1$ and $X^2$ (in the case of $T^2$)
\begin{eqnarray*}
(X^1,X^2)= (X^1+2\pi\sqrt{\alpha'}q_{(1)}s, X^2+2\pi\sqrt{\alpha'}q_{(1)}s)
\end{eqnarray*}
while
\begin{eqnarray*}
(X^1,X^2)\neq (X^1, X^2+2\pi\sqrt{\alpha'}q_{(1)}s)\qquad (X^1,X^2)\neq (X^1+2\pi\sqrt{\alpha'}q_{(1)}s, X^2)
\end{eqnarray*}
This is consistent with the fact that, with the special choice
of the T-duality transformation given in Eq. (\ref{Lambda2}),
the matrix $\D$ is purely diagonal with two identical entries.

In conclusion, the theory based on the Narain branes seems to provide
a description of branes wrapped  on the two-cycle of the torus, that is
different, rather than alternative, from that provided by the non-abelian branes. A further study of
these two different formulations of wrapped branes
is needed to better clarify their physical
properties and what kind of wrapped branes they really describe.

\appendix

\section{Conventions}
\label{app-conventions}
\begin{itemize}
\item Indices:\\
Non-compact $\mu,\nu= 0, \dots 25-\hat{d}$;

Compact $i,j,\dots= 1,\dots \hat d$;

\item $\delta^{[N]}_{m,n}$ means $m\equiv n~~~mod~N$;

\item 't Hooft matrices $P_N$ and $Q_N$:
\begin{eqnarray*}
\!\!P_N=\left( \begin{array}{cccc}
              0&1 & \dots &0\\
              \vdots &\ddots &\ddots\\
              0&   &\dots  &1\\
              1&0  &\dots  & 0
            \end{array}\right)~~;~~Q_N=e^{\frac{\pi i (1-N)}{N}}\!\!\left(\begin{array}{ccc}
                                                                       1& \dots   &0  \\
                                                                       0 &e^{\frac{2i\pi}{N}}   \dots\\
                                                                       \vdots&\ddots &\vdots\\
                                                                        0     &\dots       &e^{2i\pi\frac{(N-1)}{N}}
                                                                        \end{array}\right)
\end{eqnarray*}
satisfying the commutation relation:
\begin{eqnarray}
P_N\,Q_N=Q_N\,P_N\, e^{2\pi\,i/N}.
\label{PQ=}
\end{eqnarray}

\item Background matrices:\\
\begin{eqnarray}
E&=& \parallel E_{i j} \parallel = G+B
\nonumber\\
{\cal E} &=& \parallel \calE_{i j} \parallel=E^T + 2\pi \alpha' q_0 F
= G -{\cal B}
\label{calEc}
\end{eqnarray}
and
\begin{eqnarray*}
\hat F &=& 2\pi \alpha' q_0 F
\nonumber\\
{\cal B} &=& B - 2\pi \alpha' q_0 F = B -\hat F
\nonumber\\
{\cal E}^{-1} &=& {\cal G}^{-1} -\Theta
\end{eqnarray*}
from which we deduce that
\begin{eqnarray*}
&&{\cal E}{\cal G}^{-1} {\cal E}^T
={\cal E}^T{\cal G}^{-1} {\cal  E}
=G
\nonumber\\
&&
\Theta= \frac{1}{2}\left({\cal E}^{-T}-{\cal E}^{-1} \right)
=- {\cal E}^{-1} {\cal B} {\cal E}^{-T}
\end{eqnarray*}
\end{itemize}

\section{Review of open and closed strings in flux background}
\label{sect-reviewA}

In this Appendix we review the solution of closed string
equations of motion in  constant
backgrounds on a torus in order to fix our notations and give some technical details
about the open string solution as well.

\subsection{Action and equations of motion}
\label{bosonic}

Let us consider  the action for the spatial coordinates,  labelled
 by the indices $a,b=1, \dots,  {d}-1$, of  a bosonic
 string~\footnote{Although we consider the bosonic string where $d=26$, we leave
 $d$ arbitrary because many of the results are also valid for the superstring where
 $d=10$.}
interacting with a
constant gravitational and a Kalb-Ramond  background that is given by Eq. (\ref{acti853}).

Constant gravitational and Kalb-Ramond fields naturally arise when
considering string theory on a $\hat d$-dimensional torus
$T^{\hat d}$~\footnote{We assume that ${\hat{d}}$  spatial coordinates are
compact, while the remaining $d-1- {\hat{d}}$ are non-compact.} .
Toroidal compactification requires the following equivalence relation
to be satisfied by any point $x^{i}$ ($i=1,\dots, {\hat d}$) of the
torus $T^{{\hat d}}$:
\begin{eqnarray}
 x^{i}  \equiv {x}^i + 2 \pi \sqrt{\alpha'}m^{i}
\label{tor}
\end{eqnarray}
where $m^{i}$ is an  arbitrary integer. This relation
has to be satisfied also by the string coordinates themselves:
 \begin{eqnarray}
X^i \equiv X^i + 2 \pi \sqrt{\alpha'}m^i.
\label{mI39}
\end{eqnarray}

The classical equation of motion for the string coordinates derived
from $S$ is given
by the usual free two-dimensional wave-equation:
\begin{eqnarray}
\partial_{\alpha} \partial^{\alpha} X^j =0.
\label{equa34}
\end{eqnarray}
In order that the action be stationary under the general
variation $X^{i} \rightarrow \delta X^{i}$
we must also impose either the closed string boundary condition
\begin{equation}
X^{i}(\tau , \sigma + \pi) \equiv X^{i}(\tau, \sigma)
\label{bou25clo}
\end{equation}
or one of the two boundary conditions at $\sigma=0$:
\begin{eqnarray}
X^i |_{\sigma=0} =const
\nonumber\\
G_{ij} \partial_{\sigma} X^{j} + B_{ij} \partial_{\tau}X^{j}
|_{\sigma=0} =0.
\label{bou25}
\end{eqnarray}
and similarly, and independently, at $\sigma=\pi$.

In the presence of such non trivial backgrounds the string
conjugate momentum density turns out to be:
\begin{eqnarray}
P_i \equiv \frac{\partial L}{\partial {\dot{X}}^i}= \frac{1}{2 \pi \alpha'
  } \left[ G_{ij} {\dot{X}}^j + B_{ij} X^{'j} \right]~~\Rightarrow~~
{\dot{X}}^i = 2 \pi  \alpha' G^{ij}P_j - G^{ik} B_{kj} (X')^{j}
\label{conmo59}
\end{eqnarray}
and the Hamiltonian is given by:
\begin{eqnarray}
H = \int_{0}^{\pi} d \sigma \left[ P_i (\dot{X})^i - L \right]=
\frac{1}{4\pi \alpha'} \int_{0}^{\pi} d\sigma G_{ij}
\left( \dot{X}^{i} \dot{X}^{j} + X^{'i} X^{'j} \right).
\label{hamilto}
\end{eqnarray}
By plugging Eq. (\ref{conmo59}) in Eq. (\ref{hamilto}) we get:
\begin{eqnarray}
H = \pi \int_{0}^{\pi} d \sigma \left[ \alpha'  P_i G^{ij} P_j +
\frac{1}{\pi }
  (X')^i B_{ij} G^{jk} P_k + \frac{1}{(2\pi)^2 \alpha'} (X')^{i} \left( G_{ij}
    - B_{ik} G^{kh} B_{hj}\right) (X')^j  \right].
\label{hamilto1}
\end{eqnarray}

\subsection{General solution for the closed string}
\label{B2}

The general solution of (\ref{equa34}) compatible with the
closed-string boundary condition (\ref{bou25clo})
 is\footnote{
With respect to the notation used in
\cite{DiVecchia:2006gg} we have exchd $\alpha
\leftrightarrow \tilde \alpha$.}:
\begin{eqnarray}
X^i(\sigma,\tau) = \frac{1}{2} \left( X^{i}_{L} (\tau + \sigma ) +
\tilde X^{i}_{R} ( \tau -\sigma )
 \right)
\label{move45}
\end{eqnarray}
where  the left and right moving parts are defined as follows:
\begin{eqnarray}
X^{i}_{L} (\tau + \sigma )
=  x^{i}_{L}
+ 4\alpha' G^{ij} p_{Lj} (\tau + \sigma)
+
i \sqrt{2 \alpha'} \sum_{n \neq 0}
\frac{1}{n}\alphant_{n}^{i}  e^{-2 i n (\tau + \sigma)},
\label{xi75}
\end{eqnarray}
\begin{eqnarray}
\tilde X^{i}_{R} ( \tau -\sigma )
=
 x^{i}_{R}
+ 4 \alpha' G^{ij}p_{Rj} (\tau - \sigma)
+
i \sqrt{2 \alpha' } \sum_{n \neq 0} \frac{1}{n}
\alphat_{n}^{i} e^{-2 i n (\tau - \sigma)}.
\label{xi76}
\end{eqnarray}
One has:
\begin{eqnarray*}
x^i=x^i_L=x^i_R~~~;
~~~~
\frac{p_i}{2}=p_{Li}=p_{Ri}
\end{eqnarray*}
in non-compact directions
and
\begin{eqnarray}
p_{Li}
&=&
\frac{1}{2\sqrt{\alpha'}}
\left[  \sqrt{\alpha'} p_i -  B_{i j} m^j   + G_{ij}m^j  \right]
\nonumber\\
p_{Ri}
&=&
\frac{1}{2\sqrt{\alpha'}}
\left[\sqrt{\alpha'}p_i -  B_{j j} m^j   - G_{ij} m^j  \right]~
\label{compact4}
\end{eqnarray}
in compact directions, where $m^i\in\Z$ is the winding number. We
can invert those relations getting
\begin{equation*}
m^i= \sqrt{\alpha'} G^{ij}( p_{Li} -p_{Ri} )
~~~~
p_i=  E_{i j}G^{jk} p_{Lk} + (E^T)_{i j}G^{ik} p_{Rk}
\end{equation*}
where we have defined:
\begin{equation*}
E_{i j} \equiv G_{i j} + B_{i j}.
\end{equation*}
By expressing  the
conjugate momentum in Eq. (\ref{conmo59}) in terms
of the oscillators one gets:
\begin{eqnarray}
P_i &=& \frac{p_i}{\pi} +
\frac{1}{ \pi \sqrt{2 \alpha'}} \sum_{n \neq 0}
\left[(G_{i j} - B_{i j}) \alphat^{j}_{n} {e}^{-2 i n( \tau- \sigma)} +
(G_{i j} + B_{i j}) \alphant^{j}_{n} {e}^{-2 i n( \tau +
    \sigma)} \right]
\nonumber\\
&=&
\frac{1}{2\pi \alpha'} \left[ E_{i j} \partial_+ X^i
+(E^T)_{i j} \partial_- X^i\right]
\label{pi12}
\end{eqnarray}
where also the following relation has been used:
\begin{eqnarray}
\frac{ \partial  X^i}{\partial \sigma} &=&
2 m^i \sqrt{\alpha'}+ \sqrt{2\alpha'} \sum_{n \neq 0}
\left[ - \alphat^{i}_{n} {e}^{-2 i n( \tau- \sigma) }+
\alphant^{i}_{n} {e}^{-2in( \tau + \sigma)} \right]
\nonumber\\
&=&
\partial_+ X^i -\partial_- X^i .
\label{xprimo}
\end{eqnarray}
Of course, along the non-compact directions one has to set
$m^i\equiv 0$.

The quantization of the theory is obtained  by imposing the
following commutation relations:
\begin{eqnarray*}
[ x^i_L, p_{Lj}] = i G^{i}_{\,\, j}
~~&&~~
[ x^i_R, p_{Rj}] = i G^{i}_{\,\, j}
\nonumber\\
~ [ \alpha^i_m , \alpha^j_n]= m G^{i j} \delta_{n+m,0}
~~&&~~
[ \tilde \alpha^i_m, \tilde \alpha^j_n]= m G^{i j} \delta_{n+m,0}
\end{eqnarray*}
Those for the non-zero modes follow from imposing the
canonical commutation relations, while those involving the
zero modes are a consequence of the canonical commutation
relations and of  T-duality that requires to consider operators
both $x_{L,R}$ and  $p_{L,R}$ and not only the combinations
$x_R + x_L$ and $p_R +p_L$.

In a compact space, like $T^{\hat d}$, the total momentum
$p_i=\int_0^\pi d\sigma~ P_{i}$ has to be quantized
since all physical states must be translational invariant under the
shift in Eq. (\ref{tor}), hence for all compact directions $i$ one has:
\begin{eqnarray}
\sqrt{\alpha'}p_i = {n_i}\in\Z.
\label{inte50}
\end{eqnarray}

By inserting  the expansions in
terms of the oscillators in the Hamiltonian (\ref{hamilto1})
one gets that the spectrum is given by the following
quantity~\footnote{From now on we consider the quantity $\frac{H}{2}$
instead of just $H$ because it is this quantity that determines the
spectrum of the theory with the correct normalization.}:
\begin{equation}
\frac{H}{2}
=
\frac{1}{2}Z^T M Z
+ \frac{1}{2}
\sum_{n >0} G_{ij}
:
\left[\alpha^{i}_{-n}  \alpha^{j}_{n}
+ \alpha^{i}_{n}  \alpha^{j}_{-n}
+ {\tilde{\alpha}}^{i}_{-n} {\tilde{\alpha}}^{j}_{n}
+ {\tilde{\alpha}}^{i}_{n} {\tilde{\alpha}}^{j}_{-n}
\right]
:
\label{spectrum}
\end{equation}
where
\begin{equation}
Z \equiv \left( \begin{array}{c}\hat n_j \\
                             \hat m^j \end{array} \right)
~~,~~
M \equiv \left(
                         \begin{array}{cc} G^{ij} & - G^{ik}B_{kj} \\
          B_{ik} G^{kj}   & G_{ij} - B_{ik} G^{kh} B_{hj} \end{array} \right).
\label{zandm}
\end{equation}
being $\hat n_i$ and $\hat m^i$ operators.
It is also easy to see that:
\begin{eqnarray}
\frac{1}{2}Z^T M Z
=
\frac{1}{2}
\left[ \hat n_i G^{ij} \hat n_j
- 2 \hat n_i G^{ij}  B_{jk} \hat m^k
+ \hat m^i \left(G_{ij} - B_{ik} G^{kh} B_{hj} \right) \hat m^j
\right].
\label{spectrum2}
\end{eqnarray}
The Hamiltonian can also be written as follows:
\begin{eqnarray}
\frac{H}{2} =L_0 + \tilde L_0 \label{hamilto89}
\end{eqnarray}
with the explicit expressions of $L_0$ and ${\tilde{L}}_0$ given by:
\begin{eqnarray}
L_0 = \alpha' p_{Li} p_{lj} G^{ij}
+ \sum_{n=1}^{\infty} G_{i j} :\alpha_{-n}^{i}\alpha_{n}^{j}:
~~;~~
\tilde{L}_0 = \alpha' p_{Ri} p_{Rj} G^{ij}
+ \sum_{n=1}^{\infty} G_{i j} :\alphat_{-n}^{i}\alphat_{n}^{j}:
\label{L078}
\end{eqnarray}
where
\begin{eqnarray}
p_{\left(\begin{array}{c} Li \\
                 Ri \end{array} \right) }
&=& \frac{1}{2\sqrt{\alpha'}} \left[ \hat n_i - B_{i j} \hat m^j
\pm G_{i j} \hat m^j \right] . \label{pLpR89}
\end{eqnarray}
It is straightforward to check that the level matching condition is given by;
\begin{eqnarray}
{\tilde{L}}_0 - L_0 = \hat n_i \hat m^i + \sum_{n=1}^{\infty} G_{ij}
\left[\alphant_{-n}^{i} \alphant_{n}^{j} -
\alphat_{-n}^{i} \alphat_{n}^{j}  \right] =0.
\label{levelma}
\end{eqnarray}
The vacuum is then defined as the state satisfying the following conditions
\begin{eqnarray*}
p^{i}_{L}|0,\tilde 0\rangle
=
p^{i}_{R}|0,\tilde 0\rangle
=
\alpha^i_n|0,\tilde 0\rangle
=
\tilde \alpha^i_n|0,\tilde 0\rangle
=0
~~~~
\forall n>0
\end{eqnarray*}
The momentum states are normalized as
\begin{equation*}
\langle n_i, m^i | n'_i, m'^i \rangle = 2 \pi \sqrt{\alpha'}
\delta_{ n_i, n'_i} \delta_{m^i, m'^i}
\end{equation*}
for any compact direction $x^i$ and
\begin{equation*}
\langle k_\mu | k'_\mu \rangle =
2 \pi \delta(k_\mu-k'_\mu)
\end{equation*}
for any non-compact spatial direction $x^a \, ( a= \mu \neq i)$
and for the time direction
($\mu=0$).

In the following we will consider the boundary state corresponding to
a space filling brane. In this case, if one starts from
Eq. (\ref{bou0}) with the substitution
$\sigma \leftrightarrow \tau$,  one can write
the equation that the boundary state has to satisfy, namely:
\begin{eqnarray}
\left[ G_{ij} \partial_{\tau} X^j + ( B_{ij} - 2 \pi \alpha' q F_{ij} )
  \partial_{\sigma} X^j \right]_{\tau =0} | B \rangle =0.
\label{bou0bs}
\end{eqnarray}

 In Eq. (\ref{bou0bs}) one has to insert the general solution of the classical equations of motion (\ref{equa34})
compatible with the
closed string boundary condition $X^{i}(\tau , \sigma + \pi) \equiv X^{i}(\tau, \sigma)$.  Such solution is given in
Eq. (\ref{move45}).

In doing that  one gets the following conditions
\begin{eqnarray}
({\hat{n}}_i - 2 \pi \alpha' q F_{ij} {\hat{m}}^j ) | B \rangle =0
\label{zeromo}
\end{eqnarray}
and
\begin{eqnarray}
\left[ \left( G_{ij} - B_{ij} + 2 \pi \alpha' q F_{ij} \right)
{\tilde{\alpha}}_{n}^{j} + \left( G_{ij} + B_{ij} - 2 \pi \alpha' q F_{ij} \right)
{\alpha}_{-n}^{j} \right] | B \rangle =0
\label{othemo}
\end{eqnarray}
that can also be written as follows (by using
Eq. (\ref{inte42})):
\begin{eqnarray}
\left( {\cal{E}}_{ij} {\tilde{\alpha}}_{n}^{j} + {\cal{E}}^{T}_{ij}
{\alpha}^{j}_{-n} \right) |B \rangle =0 ,
\label{othemo2}
\end{eqnarray}
being $qF=q_0F_0$ on the boundary at $\sigma=0$ and $qF=q_\pi
F_\pi$ on the boundary at  $\sigma=\pi$.

It is easy to  rewrite Eq. (\ref{zeromo}) as follows:
\begin{eqnarray}
\left[ {\cal{E}}_{ij} p_{R}^{j} + {\cal{E}}^{T}_{ij}
  p_{L}^{j} \right] |B \rangle=0
\label{illu}
\end{eqnarray}
where ${\cal{E}}_{ij} $ is defined in Eq. (\ref{calE}) and $p^{i}_{L,R}= G^{ij} p_{j;L,R}$.

For later use here we give the explicit form of
the boundary state for a D25  brane that satisfies
Eq.s (\ref{othemo2}) and (\ref{illu}) with $F_{ij} =0$
$ (T_p =\frac{ \sqrt{\pi} }{ 2^{ \frac{d-10}{4}} }
( 2 \pi \sqrt{\alpha'} )^{ \frac{d}{2} -2 - p} )$ :
\begin{eqnarray}
 | D25 \rangle
&=& \frac{T_{25}}{2} e^{-\sum_{n=1}^\infty \frac{1}{n}
~\alpha^{t~i}_{-n}  (E^{ T})_{i k}(E^{-1})^{kh} G_{hj}
\alphat^{t~j}_{-n} } ~| D25 \rangle _{z m, c}
\nonumber\\
&\times& e^{-\sum_{n=1}^\infty \frac{1}{n} ~\alpha^{t~0}_{-n} G_{0
0}  \alphat^{t~0}_{-n} } ~| k_0=0 \rangle
\label{plain-D25a}
\end{eqnarray}
where the time direction has been added and the compact zero modes part is
given by:
\begin{eqnarray}
| D25 \rangle _{z m, c} = \frac{\sqrt{\det {
E}}}{\left(\det G\right)^{1/4}}~ \sum_{s\in Z^{25}} | n_i=0,
m^i = s^i \rangle .
\label{D25zmta}
\end{eqnarray}
In the next sections
we will include the
dependence on $F_{ij}$ on the
boundary state of a space filling brane.

\subsection{General solution for open strings: some technical details}
\label{dettagli}
In this section we solve the equation of motion and the boundary conditions in Eq. (\ref{bou0})
for an open string. To this purpose  it is convenient to rewrite
Eq. (\ref{bou0}) as follows:
\begin{eqnarray}
\left[ {\cal E}^T_{(0) i j} \partial_{+} X^j
  - {\cal E}_{(0) i j} \partial_{-} X^j \right]_{\sigma =0} =0
\label{bou02}
\end{eqnarray}
and
\begin{eqnarray}
\left[ {\cal E}^T_{(\pi) i j} \partial_{+} X^j
  - {\cal E}_{(\pi) i j} \partial_{-} X^j \right]_{\sigma =\pi} =0
\label{boupi2}
\end{eqnarray}
where we have defined
\begin{eqnarray}
{\cal{E}}_{(0) ij}
\equiv  G_{ij} - (B_{ij} - 2\pi \alpha' q_0  F_{ij}^{(0)})
= G_{ij} - \calB_{ (0) i j}
= (E^T)_{i j} +\hat F_{i j}
\label{inte42}
\end{eqnarray}
with
\begin{eqnarray}
{\cal{B}}_{(0) i j}
&\equiv&
 B_{i j} - 2 \pi \alpha' q_0 F_{i j}^{(0)}
=
 B_{i j} -  \hat F_{i j}^{(0)}
\nonumber\\
\hat F_{i j}^{(0)} &\equiv&  2 \pi \alpha' q_0 F_{i j}^{(0)}
\label{def73}
\end{eqnarray}
and similarly for the $(\pi)$ quantities.

The general solution of the bulk equation in (\ref{bulkequa}) is given
by:
\begin{eqnarray}
X^{i} ( \sigma, \tau) =
\frac{1}{2} \left[
G^{i j} {\cal E}_{(0) j k} F^{k} (\tau + \sigma)
+G^{i j} ({\cal E}^T)_{(0) j k} G^{k} (\tau - \sigma)
\right]
\label{solu83}
\end{eqnarray}
with $F^{i} (\tau + \sigma)$  and $G^{i} ( \tau - \sigma )$ arbitrary
functions.

We have chosen the particular form in Eq. (\ref{solu83}) because it immediately
solves the  boundary condition at $\sigma =0$  as we will show shortly.
By inserting Eq. (\ref{solu83})  in the two boundary conditions one gets:
\begin{eqnarray}
({\cal E}^T_{(0)} G^{-1} {\cal E}_{(0)})_{i j} \partial_{\tau} F^{j} (\tau)
=
({\cal E}_{(0)} G^{-1} {\cal E}^T_{(0)})_{i j} \partial_{\tau} G^{j} (\tau)
\label{boun42}
\end{eqnarray}
and
\begin{eqnarray}
({\cal E}^T_{(\pi)} G^{-1} {\cal E}_{(0)})_{i j} \partial_{\tau} F^{j} (\tau +  \pi)
=
({\cal E}_{(\pi)} G^{-1} {\cal E}^T_{(0)})_{i j}
\partial_{\tau} G^{j} (\tau -\pi) .
\label{boun43}
\end{eqnarray}
Let us remind here that $G^{ij}$ means the inverse of the matrix
$G_{ij}$, i.e. $G^{ik}G_{kj}=\delta^{i}_{j}$. In the following we will denote $G^{ij}$
with $G^{-1}$ only when the indices $i$ and $j$ are not explicitly written. We are
also using this convention for all other matrices.
The  boundary condition  at $\sigma=0$ is immediately solved by
\begin{equation}
G^i(\tau)= F^i(\tau) + const
\label{G-F-const}
\end{equation}
since the open string metric $\calG_{(0)}$
satisfies the relation:
\begin{equation*}
\calG_{(0)}
= \calE_{(0)}^T G^{-1} \calE_{(0)}
= \calE_{(0)} G^{-1} \calE_{(0)}^T.
\end{equation*}
In order to solve the  boundary condition at $\sigma = \pi$
it is convenient to
introduce the quantity:
\begin{eqnarray}
R ^{i}_{\,\,j}
=
\left( ( \calE^T_{(\pi)} G^{-1} \calE_{(0)} )^{-1} \right)^{i k}
\left( \calE_{(\pi)} G^{-1} \calE^T_{(0)} \right)_{k j}
=
\left( \calE^{-1}_{(0)} G \calE^{-T}_{(\pi)}
        \calE_{(\pi)} G^{-1} \calE^T_{(0)} \right)^{i}_{\,\, j}
\label{R0pi}
\end{eqnarray}
which is a $SO(\hat d)$ matrix with respect to the metric $\calG_{(0)}$
\begin{equation*}
R^T \calG_{(0)} R = \calG_{(0)}.
\end{equation*}
The  boundary condition at $\sigma=\pi$ can now be
written as follows:
\begin{eqnarray}
\partial_{\tau} F^{i} (\tau + \pi) =
R^{i}_{\,\,j} \partial_{\tau} F^{j} (\tau - \pi).
\label{boun31}
\end{eqnarray}
In order to solve the previous equation one should diagonalize the $R$-matrix. However,
in the dipole case, one can avoid such a problem because:
\begin{equation*}
q_{0}F^{(0)} - q_{\pi}F^{(\pi)}=0 \Rightarrow R=\uno.
\end{equation*}
In this case all the $(0)$ quantities drop and one can simply write
$\calE$ for $\calE_{(0)}$ and so on.
The solution of Eq. (\ref{boun31}) is:
\begin{eqnarray}
\partial_{\tau} F^{i} ( \tau + \sigma)
= \sqrt{ 2 \alpha'}
\sum_{n =  -\infty}^{\infty}
\left(
     {e}^{ - i ( \tau+ \sigma )
      n \uno }
  \right)^{i}_{\,\,j} \alpha^{j}_{n}
\label{solu56}
\end{eqnarray}
that  can be integrated to give:
\begin{eqnarray}
F^{i} ( \tau + \sigma)
=
x^i
+ i \sqrt{2 \alpha' }
 \sum_{n = -\infty}^{\infty}
\left( \frac{1}{n}
       {e}^{ - i ( \tau+ \sigma )
        n \uno}
\right)^{i}_{\,\,j}
\alpha^{j}_{n}
\label{solu57}
\end{eqnarray}
where $x^i$ is an arbitrary constant of integration.
The
open string
expansion (excluding pure Dirichlet boundary conditions)\footnote{
With respect to the conventions used in \cite{Chu:2000wp} (CRS) and in
\cite{Chu:2005ev} (C) we have
$  G=g_{C S R}=g_{C}, {\cal B}=-F_{C R S}=2\pi \alpha' {\cal B}_{C},
{\cal G}=M_{C R S} = G_{C},
 2\pi \alpha' \Theta=\Theta_{C S R}=\Theta_{C}$.
}:\\
\begin{eqnarray}
X^i(\sigma,\tau) &=&
\frac{1}{2}\left(\hat X_L^i(\tau+\sigma)+ \hat X_R^i(\tau-\sigma) \right)
\label{icsoa}
\end{eqnarray}
where $\hat X_L^i(\tau+\sigma)$ and $\hat X_R^i(\tau-\sigma)$ are the ones already written respectively in (\ref{icsoaa}) and (\ref{icsob}).
It is also useful to define the commuting coordinates $x_0^i$
\begin{eqnarray}
x^i= x_0^i - \pi\alpha' \Theta^{ i j} {\cal G}_{j k} p^k
\label{cofa}
\end{eqnarray}
where $x_0$ satisfies the usual commutation relations
\begin{eqnarray}
[x^i_0,x^j_0]=0
~~;~~
[x^i_0,p^j]=i \calG^{i j}\label{comzm}
\end{eqnarray}
Given the operator $x_0$, we define:
\begin{eqnarray}
\label{icsol}
\hat X_L^i(\tau+\sigma)
&=&
\hat X_{L (0)}^i(\tau+\sigma)
+ \pi\alpha' (G^{-1} {\cal E}  \Theta {\cal G})^i_j p^j
\nonumber\\
&=&
(G^{-1} {\cal E})^i_j
\left( X_{L (0)}^j(\tau+\sigma)
 + \pi\alpha' (\Theta {\cal G})^j_l p^l \right)
\nonumber\\
&=&
(G^{-1} {\cal E})^i_j
\left( X_{L (0)}(\tau + \sigma) + \pi\alpha'
G^{-1} {\cal B} p \right)^j
\end{eqnarray}
and
\begin{eqnarray}
\label{icsor}
\hat X_R^i(\tau-\sigma)
&=&
\hat X_{R (0)}^i(\tau-\sigma)
+\pi\alpha' (G^{-1} {\cal E}^T   \Theta {\cal G})^i_j p^j
\nonumber\\
&=&
(G^{-1} {\cal E}^T)^i_j
\left( X_{R (0)}^j(\tau-\sigma)
+\pi\alpha' (\Theta {\cal G})^i_j p^j \right)
\nonumber\\
&=&
(G^{-1} {\cal E}^T)^i_j
\left( X_{R(0)}(\tau-\sigma)
+ \pi \alpha' G^{-1} {\cal B} p \right)^j
\end{eqnarray}
where all the quantities with $(0)$ depend on $x_0$ instead of $x$.
Here we have introduced
\begin{eqnarray*}
 X_{L (0)}(z)
&=&
x_0 -2\alpha' i p \ln z
+ i \sqrt{2\alpha'} \sum_{n\ne 0} \frac{sgn(n)}{\sqrt{|n|}} a_n z^{-n}
~~~~
0\le arg(z) \le \pi
\nonumber\\
 X_{R(0)}(\bar z)
&=&
x_0 -2\alpha' i p \ln \bar z
+ i \sqrt{2\alpha'} \sum_{n\ne 0} \frac{sgn(n)}{\sqrt{|n|}} a_n \bar z^{-n}
~
-\pi \le arg(\bar z) \le 0
\end{eqnarray*}
where $z=e^{i(\tau+\sigma)}$,
Notice that for $\sigma=0$ equations (\ref{icsoa}),(\ref{icsol}) and (\ref{icsor})
becomes
\begin{eqnarray}
X^i(\tau) &=&
\frac{1}{2}\left(\hat X_L^i(\tau)+ \hat X_R^i(\tau) \right)
\nonumber\\
&=& X_{L(0)}(\tau)-\pi\alpha' (\Theta{\cal G})^i_jp^j
\label{icsot}
\end{eqnarray}
where we have used that $X_{L(0)}(x)=X_{R(0)}(x)$.

The spectrum of $p^i$ is given by
\begin{equation*}
\calG_{i j} p^j |k\rangle = k_i |k \rangle
= \frac{n_i}{\sqrt{\alpha'}} -q_\pi a_{(\pi) i} +q_0 a_{(0) i} |k \rangle
\end{equation*}
with $n_i\in\Z$ and
where $a^{(0,\pi)}_i$ are the constant parts of the gauge fields
$A^{(0,\pi)}_i$.

The OPEs read
\begin{eqnarray}
X_{L (0)}(z) X^{T}_{L (0)}(w) &=& -2 \alpha' \ln(z-w) {\cal G}^{-1}
\nonumber\\
X_{L (0)}(z) X^{T}_{R (0)}(\bar w) &=& -2 \alpha' \ln(z-\bar w) {\cal G}^{-1}
\nonumber\\
X_{R (0)}(\bar z) X^{T}_{R (0)}(\bar w) &=& -2 \alpha' \ln(\bar z-\bar
w) {\cal G}^{-1}
\label{openOPEs-X0}
\end{eqnarray}
or using ${\cal E}{\cal G}^{-1} {\cal E}^T
={\cal E}^T{\cal G}^{-1} {\cal  E} =G$
\begin{eqnarray}
\hat X_{L }(z) \hat X^{T}_{L }(w) &=& -2 \alpha' \ln(z-w) {G}^{-1}
\nonumber\\
\hat X_{L }(z) \hat X^{T}_{R }(\bar w)
&=&
-2 \alpha' \ln(z-\bar w)
G^{-1} {\cal E} {\cal G}^{-1} {\cal E} G^{-1}
=
-2 \alpha' \ln(z-\bar w)
{\cal E}^{-T} {\cal E} G^{-1}
\nonumber\\
\hat X_{R }(\bar z) \hat X^{T}_{R }(\bar w) &=& -2 \alpha' \ln(\bar z-\bar
w) {G}^{-1}
\label{openOPEs-X}
\end{eqnarray}

\section{Short review of closed string canonical linear transformations}
\label{T-duality}

A general T-duality transformation is a canonical
transformation of the form
\begin{eqnarray}
\left( \begin{array}{c} \frac{X'^t}{2\pi \alpha'} \\ { P}^t
\end{array}\right) &=&\Lambda \left( \begin{array}{c} \frac{X'}{2\pi
\alpha'} \\ { P} \end{array}\right) \label{CanTras}
\end{eqnarray}
{\rm with}
\begin{eqnarray}
\Lambda &=& \left( \begin{array}{cc} \A & \B \\ \C & \D
\end{array}\right) \in O(\hat{d},\hat{d},Z)
\label{lam}
\end{eqnarray}
where $X=\parallel X^i\parallel$ and $P=\parallel P_i\parallel$ with
$i=1,\dots {\hat d}$ are column vectors.
Here and in what follows, the momentum $p$ is understood with covariant indices, unless explicitly indicated.
To belong to the group $O(\hat{d},\hat{d},Z)$ the matrix $\Lambda$ must be a
${\hat d}\times {\hat d}$ matrix with integer entries satisfying the constraint
\begin{equation}
\Lambda \left( \begin{array}{cc} 0 & 1_{\hat{d}} \\ 1_{\hat{d}} & 0
\end{array}\right) \Lambda^T = \left( \begin{array}{cc} 0 & 1_{\hat{d}} . \\
1_{\hat{d}} & 0 \end{array}\right)\equiv J \label{CanTransMatrix}
\end{equation}
This constraint simply follows from the canonical commutation
relations:
\begin{equation*}
[ X'(\sigma), { P}^T(\sigma')]= [ { P}(\sigma), { X'}^T(\sigma')]= i
\partial_\sigma \delta(\sigma-\sigma') 1_{\hat{d}}
\end{equation*}
which imply
\begin{equation}
\left[ \left( \begin{array}{c} \frac{X'}{2\pi \alpha'} \\ { P}
\end{array}\right)(\sigma), \left( \frac{X'^T}{2\pi \alpha'} , {
P}^T\right)(\sigma') \right]= \left( \begin{array}{cc} 0 & 1_{\hat{d}} \\ 1_{\hat{d}} & 0
\end{array}\right) \frac{i}{2\pi \alpha'} \partial_\sigma
\delta(\sigma-\sigma'). \label{comm1}
\end{equation}
Under the transformation in Eq. (\ref{CanTras}) the previous
commutator becomes
\begin{eqnarray}
\left[ \left( \begin{array}{c} \frac{X'^t}{2\pi \alpha'} \\ { P^t}
\end{array}\right)(\sigma), \left( \frac{(X'^t)^T}{2\pi \alpha'} , {
P^t}^T\right)(\sigma') \right]=\Lambda\left( \begin{array}{cc} 0 & 1_{\hat{d}} \\
1_{\hat{d}} & 0
\end{array}\right)\Lambda^T \frac{i}{2\pi \alpha'} \partial_\sigma
\delta(\sigma-\sigma') \label{comm2}
\end{eqnarray}
However  the commutation relation in (\ref{comm1}) has to be invariant
under the transformation in Eq. (\ref{CanTras}) and thus equating
the left hand sides of Eq.s (\ref{comm1}) and  (\ref{comm2}) one gets
Eq. (\ref{CanTransMatrix}) implying  $\Lambda\in O(\hat{d},\hat{d},R)$. In order to
derive the constraint $\Lambda\in O(\hat{d},\hat{d},Z)$ for the T-duality group
we have to work a little more and we first define:
\begin{eqnarray*}
\partial\equiv \frac{\partial}{\partial(\sigma+\tau)}
\qquad\bar{\partial}\equiv \frac{\partial}{\partial(\tau-\sigma)}
\end{eqnarray*}
and then we get:
\begin{equation*}
X' = \partial X_L(\tau+\sigma) - \bar \partial
\tilde{X}_R(\tau-\sigma) ~~~~ {2\pi \alpha'}{ P} = E
\partial {X}_L + E^T \bar \partial \tilde{X}_R
\end{equation*}
with $E_{ij}=G_{ij}+B_{ij}$, being $X_L$ and $\tilde{X}_R$ defined
in Eq.s (\ref{xi75}) and (\ref{xi76}). Eq.s (\ref{CanTras}) can
be split into an holomorphic and an antiholomorphic part as
\begin{eqnarray}
\left(\begin{array}{c} \partial {X}^t_L \\ E^{t } \partial {X}^t_L
\end{array}\right) =\Lambda \left( \begin{array}{c} \partial {X}_L  \\ E
\partial {X}_L \end{array}\right) ~~~~~
\left(\begin{array}{c}-\bar\partial \tilde{X}_R^t \\ E^{t
T}\bar\partial \tilde{X}_R^t
\end{array}\right) =\Lambda \left( \begin{array}{c}-\bar\partial
\tilde{X}_R
\\ E^T \bar\partial \tilde{X}_R \end{array}\right) \label{hol-antihol-trans}
\end{eqnarray}
Looking first at the zero modes we get the following equations
for the left and right momenta
\begin{equation}
\left\{\begin{array}{cc}
(p^t_L)^i= & \left( \A + \B E \right)^i_{\,j} p_L^j \\
(E^{t }p^t_L)_i= & \left( \C + \D E \right)_{ij} p_L^j \\
\end{array}
\right. ~~~~ \left\{\begin{array}{cc}
(p^t_R)^i= & \left(\A - \B E^T \right)^i_{\,j} p_R^j \\
(E^{t T}p^t_R)_i= & \left( -\C + \D E^T \right)_{ij} p_R^j \\
\end{array}
\label{p-pr-transf} \right.
\end{equation}
and, if we remember that windings and momenta in compact space are
defined as
\begin{equation}
m= \sqrt{\alpha'}G^{-1}( p_L -p_R ) ~~~~ n= \sqrt{\alpha'}( E G^{-1}p_L + E^TG^{-1}
p_R ) \label{emmep}
\end{equation}
where $m=\parallel m^i\parallel$ and $n=\parallel n_i\parallel$ and the momenta are understood with covariant indeces, we
easily get that under the transformation in Eq. (\ref{p-pr-transf})
$m$ and $n$ transform as
\begin{equation}
\left\{\begin{array}{cc}
m^t = & \A m + \B n \\
n^t = & \C m + \D n \\
\end{array}
\right. \leftrightarrow \left(\begin{array}{c} m^t \\ n^t
\end{array}\right) =\Lambda \left(\begin{array}{c} m \\ n
\end{array}\right) \label{mtnt-mn}
\end{equation}
which implies the desired constraint, i.e. $\Lambda\in O(\hat{d},\,\hat{d},\,
{Z})$.

If we now consider the other terms we get the following equations
for the left and right oscillators
\begin{equation}
\left\{\begin{array}{cc}
\alpha_n^t= & \left( \A + \B E \right) \alpha_n \\
\tilde \alpha_n^t= & \left( \A - \B E^T \right) \tilde \alpha_n
\end{array}
\right. ~~~ n\in Z^* \label{at-a-closed}
\end{equation}
For the sake of completeness we collect here some consequences of Eq.
(\ref{CanTransMatrix}). We find the expression for the inverse
transformation matrices $\Lambda^{-1}$  and $\Lambda^{-T}$ to be
\begin{eqnarray}
\Lambda^{-1} &=& \left( \begin{array}{cc} 0 & 1_{\hat{d}} \\ 1_{\hat{d}} & 0
\end{array}\right) \Lambda^T \left( \begin{array}{cc} 0 & 1_{\hat{d}} \\ 1_{\hat{d}}
& 0 \end{array}\right) = \left( \begin{array}{cc} \D^T & \B^T \\
\C^T & \A^T \end{array}\right) \label{Lambda-1}\\
\Lambda^{-T} &=&\left( \begin{array}{cc} 0 & 1_{\hat{d}} \\ 1_{\hat{d}} & 0
\end{array}\right) \Lambda\left( \begin{array}{cc} 0 & 1_{\hat{d}} \\ 1_{\hat{d}}
& 0 \end{array}\right) = \left( \begin{array}{cc} \D & \C \\
\B & \A \end{array}\right) \label{Lambda-T}
\end{eqnarray}
so that Eq.  (\ref{CanTransMatrix})
can be explicitly written  as
\begin{eqnarray}
 \Lambda\, J\, \Lambda^T&=& \left(\begin{array}{cc}
\B \A^T + \A  \B^T & \B \C^T+\A \D^T\\
\D \A^T +\C \B^T & \D \C^T+ \C \D^T
\end{array}
\right)= \left( \begin{array}{cc} 0 & 1_{\hat{d}} \\ 1_{\hat{d}} & 0
\end{array}\right) \label{LLT}
\\
\Lambda^T\,J\,\Lambda &=& \left(\begin{array}{cc}
\A^T \C + \C^T \A & \A^T \D + \C^T \B\\
\B^T \C + \D^T \A & \B^T \D + \D^T \B
\end{array}
\right) = \left( \begin{array}{cc} 0 & 1_{\hat{d}} \\ 1_{\hat{d}} & 0
\end{array}\right) \label{LTL}
\end{eqnarray}
where in the second equality we have used the following identity:
\begin{eqnarray*}
\Lambda \, J\,\Lambda^T=J \Rightarrow \Lambda\, J\,
\Lambda^T\,J=\mathbb{I}\Rightarrow \Lambda\, J\, \Lambda^T\,
J\,\Lambda =\Lambda\Rightarrow  \Lambda^T\, J\, \Lambda =J
\end{eqnarray*}

The inverses of Eq.s (\ref{mtnt-mn}) and (\ref{at-a-closed}) can then
be obtained using Eq. (\ref{Lambda-1}), i.e $\A\leftrightarrow \D^T,
\B\rightarrow \B^T, \C \rightarrow \C^T$ and exchanging $^t$
quantities with those without $^t$, explicitly
\begin{eqnarray}
& \left\{\begin{array}{cc}
m = & \D^T m^t + \B^T n^t \\
n = & \C^T m^t + \A^T n^t \\
\end{array}
\right.
\nonumber\\
& \left\{\begin{array}{cc}
\alpha_n= & \left( \D^T + \B^T E^t \right) \alpha_n^t \\
\tilde \alpha_n= & \left( \D^T - \B^T E^{t T} \right) \tilde
\alpha_n^t
\end{array}
\right. ~~~ n\in Z^* \label{inverse-mn-a}
\end{eqnarray}

{From} the two Eq.s (\ref{hol-antihol-trans}) we get two different
expressions for the relation between $E^t$ and $E$
\begin{equation}
E^{t T}= (-\C + \D E^T) ( \A - \B E^T)^{-1} ~~~ E^{t }=(\C + \D E)
(\A + \B E)^{-1} \label{Et-E}
\end{equation}
which are compatible because of Eq.s (\ref{LTL}).

Finally we give the transformation properties of the background
metric
\begin{eqnarray}
G^t= (\A + \B E)^{-T}  G (\A + \B E)^{-1}=( \A - \B E^T)^{-T} G ( \A
- \B E^T)^{-1}  \label{Gt-G}
\end{eqnarray}
which can be easily shown by writing $G^t=E^t+E^{tT}$ and using
respectively the second Eq. in (\ref{Et-E}) with its transposed and
the first Equation in (\ref{Et-E}) with its transposed.

Finally it is useful to establish the connection between our
notation and the one used in \cite{9401139}. To this purpose we must
consider the closed string Hamiltonian given in Eq.(\ref{spectrum}).
By requiring the Hamiltonian to be invariant under the $O(\hat{d},\hat{d},Z)$
transformation
\begin{eqnarray*}
\frac{H^t}{2} = \frac{1}{2}(Z^t)^T M^t Z^t+\dots =
\frac{1}{2}(\Lambda Z)^T \Lambda^{-T} M \Lambda^{-1}(\Lambda
Z)+\dots\equiv \frac{H}{2}
\end{eqnarray*}
one gets $M^t=\Lambda^{-T} M \Lambda^{-1}$. By comparing such
transformation with Eq. (2.4.19) of Ref.\cite{9401139} we get that
$g=\Lambda^{-T}$. By using the expressions of $g$ as given in
Ref.\cite{9401139} and $\Lambda$ as defined in Eq.
(\ref{CanTransMatrix}), we get:
\begin{eqnarray*}
g=\left(\begin{array}{cc}
         a&b\\
         c&d
         \end{array}\right)=
         \left( \begin{array}{cc} \D & \C \\ \B & \A
         \end{array}\right)
\end{eqnarray*}
where we have used the identity $\Lambda^{-T}= J\Lambda J$, which
trivially follows from Eq. (\ref{CanTransMatrix}).
In the last part of this Appendix we give some relations useful to determine the normalization of the boundary state
given in
Eq. (\ref{D25zm}). To this aim we notice that:
\begin{eqnarray}
\frac{( \det E^t )^2}{ \det G^t}
&=& \left( \frac{  \det( -\C + \D E^T )  }{ \det (\A- \B E^T) }
\right)^2 \frac{\det (\A- \B E^T)^T \det (\A- \B E^T) }{\det G}
\nonumber\\
&=& \left\{\begin{array}{cc}
(\det (-\C + \D E^T) )^2 \frac{1}{\det G} & \det \D = 0  \\
(\det \D )^2 (\det {\cal E})^2 \frac{1}{\det  G} & \det \D \ne 0
\end{array}
\right. \label{Norm-D25zm}
\end{eqnarray}
where
we have used  the first equation in (\ref{Et-E}) and the second  in
(\ref{Gt-G}) together with the equality
\begin{equation*}
\det{\D}^{-1}  \det( -\C + \D E^T )=\det(F+E^T)=\det{\cal E}
\end{equation*}
which follows from Eq.s (\ref{effe})  (\ref{inte42}).
In this way we have determined the zero mode part
of the boundary state. Moving to the non-zero modes
we notice that
\begin{eqnarray}
\alpha_{-n}^{t T}   E^{t T}E^{t -1}G^t \tilde \alpha_{-n}^t &=&
\alpha_{-n}^{ T} (\C + \D E)^{T} (-\C+\D E^{ T})^{-T} G
\alpha_{-n}^t
\nonumber\\
&\Rightarrow& \alpha_{-n}^{ T}  {\cal E}^{ }{\cal E}^{ -T} G \tilde
\alpha_{-n}
 ~~~\det \D \ne 0
\label{B-nzm-can}
\end{eqnarray}
where we have used Eq.s (\ref{at-a-closed}) and (\ref{Et-E})
together with the following identity:
\begin{eqnarray*}
(\C+ \D E)^T(-\C+\D E^T)^{-T}= (\D^{-1}\C+ E)^T (-\D^{-1}\C +
E^T)^{-T} ={\cal E}{\cal E}^{-T}.
\end{eqnarray*}

\section{Boundary state: closed string calculation}
\label{bouclosed}
In this Appendix we determine directly in the closed string channel, the compact part of the boundary state describing a non-abelian  brane  compactified on $T^2$ and in the presence of a background gauge field with constant field strength. The generalization of such calculation to a generic torus $T^6$ will be trivial. In order to simplify  the calculation, we take  the background gauge field in the  gauge:
\begin{eqnarray*}
A_1= 0 \qquad A_2= -F_{12}x^1\label{bgf}
\end{eqnarray*}
which is different from the gauge choice made in the Sec. \ref{sect-review}. Here, the $x^i$ are the compact coordinates
bounded between $0, \,2\pi\sqrt{\alpha'}$, i.e.  $0\leq x^i<2\pi\sqrt{\alpha'}$, with $i=1,2$.

The boundary state in the presence of a magnetic field is related to the uncharged one by the relation\cite{CLNY88}:
\begin{equation}
|D25(E,F)\rangle
= Tr\left( P~e^{-i\oint q A} \right) ~
|D25(E,F=0)\rangle
\label{bsf}
\end{equation}
Denoting by $\gamma$ the closed path of the integration, we parameterized it as follows:
\begin{eqnarray*}
\gamma:~~ \sigma\in [0,\,\pi] \longrightarrow (X^1(\sigma),\,X^2(\sigma))
\end{eqnarray*}
The path, in general, will wrap  $w^{i}$ times  the torus and the details of such a wrapping are important in the evaluation of the path-ordering appearing in the Eq.(\ref{bsf}).
This is  because every time that the curve makes a turn around the cycles of the torus, the gauge transition functions must be introduced  ``to glue'' the fields at the boundaries of the torus. Such a gluing can be realized as follows. We first choose the origin of the compact frame coincident with the first end of the curve and label with  $\lambda_i$, ($\lambda_0=0$ and $\lambda_{M+1}=\pi$) the values which
the parameter $\sigma$ takes when  the path cross the boundary values
$x^i=0,\, 2\pi\sqrt{\alpha'}$, $i=1,2$.
We can write then:

\begin{eqnarray*}
X^i(\sigma)=x^i(\sigma)+ 2\pi\sqrt{\alpha'}\sum_{k=1}^p\,s_k^i
~~;~~~\sigma\in[\lambda_p,\,\lambda_{p+1}]
\end{eqnarray*}
 Here $0\leq x^i(\sigma)<2\pi\sqrt{\alpha'}$ and  $s_p^i = -1, 1$ respectively if the path in the corresponding interval $[\lambda_{p-1}, \lambda_{p}]$ ``unwraps'' or wraps once, while is zero if the curve is constant in the interval. The total wrapping will be given by:
\begin{eqnarray*}
w^i=\sum_{k=1}^M s_k^i
\end{eqnarray*}
Now, we are ready to explicitly compute the path-ordering introduced in Eq. (\ref{bsf}) in the case of non-abelian branes. It is given by:
\begin{eqnarray*}
{\rm Tr}\left( P~e^{-i\oint q A} \right)\! =\!{\rm Tr}\left[ e^{i\int_{\lambda_0}^{\lambda_1} q\,F_{12}x^1{x'}^2d\tau}\Omega_2^{s_1^2}\Omega_1^{s_1^1}\dots
 e^{i\int_{\lambda_p}^{\lambda_{p+1}} q\,F_{12}x^1{x'}^2d\tau}\Omega_2^{s_{p+1}^2}\Omega_1^{s_{p+1}^1}\dots\right]
\end{eqnarray*}
being, in this gauge, the $U(1)$ factor of the gauge transition function slightly different from the one given in Sec. \ref{sect-review}:
\begin{eqnarray*}
\Omega_1= e^{-2\pi i \sqrt{\alpha'} q\,F_{12}x^2}\omega_1 \qquad \Omega_2=\omega_2
\end{eqnarray*}
By using the previous parametrization of the curve $\gamma$, we can write:
\begin{eqnarray*}
e^{i\int_{\lambda_p}^{\lambda_{p+1}} q\,F_{12}x^1{x'}^2d\sigma}\Omega_2^{s_{p+1}^2}\Omega_1^{s_{p+1}^1 }\!\!&=&\!\!
e^{i\int_{\lambda_p}^{\lambda_{p+1}}q\, F_{12}X^1{X'}^2d\sigma-2\pi i\sqrt{\alpha'}
\sum_{k=1}^p q\,F_{12} s^1_k(X^2(\lambda_{p+1})-X^2(\lambda_p))}\nonumber\\
\!\!\!&\times&\!\!\! e^{-2\pi i \sqrt{\alpha'} s_{p+1}^1 q\,F_{12}(X^2(\lambda_{p+1})- 2\pi\sqrt{\alpha'} \sum_{k=1}^{p+1}
s^2_k)}\omega_2^{s^2_{p+1}}\omega_1^{s^1_{p+1}}
\end{eqnarray*}
which implies:
\begin{eqnarray}
P~e^{-i\oint q A} &=& e^{i\int_{\lambda_0}^{\lambda_{M+1}}q\, F_{12}X^1{X'}^2d\sigma -2\pi i\sqrt{\alpha'}\sum_{p=1}^M
\sum_{k=1}^p q\,F_{12} s^1_k(X^2(\lambda_{p+1})-X^2(\lambda_p))}\nonumber\\
&\times& e^{-2\pi i \sqrt{\alpha'} \sum_{p=0}^{M-1} s_{p+1}^1 q\,F_{12}(X^2(\lambda_{p+1})- 2\pi\sqrt{\alpha'} \sum_{k=1}^{p+1}
s^2_k) }\nonumber\\
&\times&\omega_2^{s_1^2}\,\omega_1^{s_1^1}\,\omega_2^{s_2^2}\,\omega_1^{s_2^1}\dots \omega_2^{s_M^2}\,\omega_1^{s_M^1}\label{10a}
\end{eqnarray}
The reordering of the last factor gives:
\begin{eqnarray*}
\omega_2^{s_1^2}\,\omega_1^{s_1^1}\,\omega_2^{s_2^2}\,\omega_1^{s_2^1}\dots \omega_2^{s_M^2}\,\omega_1^{s_M^1} =e^{-(2\pi \sqrt{\alpha'})^2 i F_{12}\sum_{p=1}^M s_p^1\sum_{k=1}^p s_k^2} \omega_1^{w^1}\,\omega_2^{w^2}
\end{eqnarray*}
which cancels the last factor in the second line of the Eq. (\ref{10a}), while observing that:
\begin{eqnarray*}
\sum_{p=1}^M\left[ \sum_{k=1}^p s^1_k(X^2(\lambda_{p+1})-X^2(\lambda_p))+ s^1_p X^2(\lambda_{p})\right]
= w^1\,X^2(\pi)
\end{eqnarray*}
we can write:
\begin{eqnarray*}
{\rm Tr}\left( P~e^{-i\oint q A} \right)= e^{i\int_{0}^{\pi}q\, F_{12}X^1{X'}^2d\sigma}
e^{-2\pi i \sqrt{\alpha'} F_{12}w^1\, X^2(\pi)}{\rm Tr}\left[  \omega_1^{w^1}\,\omega_2^{w^2}\right]
\end{eqnarray*}
In particular the path ordering must be
evaluated on the
string coordinate expansion with the result
\begin{equation*}
P~e^{-i\oint q A}
=
e^{i 2\pi q \left[ F_{i j}  x^i \hat m^j
    -\pi F_{i j} \hat m^i \hat  m^j\right]}
e^{- \pi \alpha' q F_{i j}
  \sum_{n=1}^\infty  (a^i_n - \tilde a^i_{-n} )
  (a^j_{-n} - \tilde a^j_{n} )
}
\omega_1^{\hat m^1 }~\omega_2^{\hat m^2 }
\end{equation*}
where now all $\hat m^i$ are winding operators.

Let us now determine the non-zero modes contribution of the boundary state, starting from
the expression  given in Eq. (\ref{plain-D25a}). The latter  corresponds to evaluate the following product of operators:
\begin{eqnarray*}
&&e^{- \pi \alpha' q F_{i j}  (a^i - \tilde a^{i\dagger} )
(a^{j\dagger} - \tilde a^j ) }
e^{-
~a^{i\dagger} G_{i h} (E^{-1})^{h k} (E^{ T})_{k j} \tilde a^{j\dagger} }
|0>
\end{eqnarray*}
which, since $ (a^i - \tilde a^{i\dagger} )$ and
$(a^{j\dagger} - \tilde a^j ) $ commute, can be easily evaluated with the introduction of an auxiliary variable $z$:
\begin{eqnarray*}
e^{- \pi \alpha' q F_{i j}  (a^i - \tilde a^{i\dagger} )
(a^{j\dagger} - \tilde a^j ) }= \int \prod_i \frac{d z_i ~d \bar z^i}{ \pi}
e^{- z_i \bar z^i
+ \pi \alpha' q F_{i j}  (a^i - \tilde a^{i\dagger} ) \bar z^j
-(a^{i\dagger} - \tilde a^i ) z_i
}
\end{eqnarray*}
The previous integral can be performed and one gets:
\begin{eqnarray*}
&&
\int \prod_i \frac{d z_i ~d \bar z^i}{ \pi}
e^{- z_i \bar z^i
+ \pi \alpha' q F_{i j}  (a^i - \tilde a^{i\dagger} ) \bar z^j
-(a^{i\dagger} - \tilde a^i ) z_i
}
~
e^{-
~a^{i\dagger} G_{i h} (E^{-1})^{h k} (E^{ T})_{k j} \tilde a^{j\dagger} }
|0>\nonumber\\
&&\hspace{1em}=
\int \prod_i \frac{d z_i ~d \bar z^i}{ \pi}
e^{- z_i \bar z^i }
e^{ \pi \alpha' q F_{i j}  a^i  \bar z^j }
e^{ -\pi \alpha' q F_{i j}  \tilde a^{i\dagger}  \bar z^j }
e^{-a^{i\dagger}  z_i }
e^{ \tilde a^i  z_i }
~
e^{-
~a^{i\dagger} G_{i h} (E^{-1})^{h k} (E^{ T})_{k j} \tilde a^{j\dagger} }
|0>
=
\nonumber\\
&&\hspace{1em}=
\int \prod_i \frac{d z_i ~d \bar z^i}{ \pi}
e^{- z_i \bar z^i }
\nonumber\\
&&\hspace{4em}
e^{ \pi \alpha' q (G^{-1})^{i n} F_{n l}   \bar z^l
\left[
-G_{i h} (E^{-1})^{h k} (E^{ T})_{k j} \tilde a^{j\dagger}
-\left(G_{i h} (E^{-1})^{h k} (E^{ T})_{k j} (G^{-1})^{m  j}
+\delta_i^j \right) z_j
\right]
}
\nonumber\\
&&\hspace{4em}
e^{
-a^{i\dagger} G_{i h} (E^{-1})^{h k} (E^{ T})_{k j} \tilde a^{j\dagger}
-a^{i\dagger} \left(G_{i h} (E^{-1})^{h k} (E^{ T})_{k m} (G^{-1})^{m  j}
+\delta_i^j
               \right)  z_j
-\pi \alpha' q F_{i j} \bar z ^j \tilde a^{i\dagger}
}
|0>
=
\nonumber\\
&&\hspace{1em}=
\int \prod_i \frac{d z_i ~d \bar z^i}{ \pi}
e^{- z_i \bar z^i }
e^{ \pi \alpha' q (G^{-1})^{i n} F_{n l}   \bar z^l
\left[
-G_{i h} (E^{-1})^{h k} (E^{ T})_{k j} \tilde a^{j\dagger}
-2 G_{i h} (E^{-1})^{h j}  z_j
\right]
}
\nonumber\\
&&\hspace{4em}
e^{
-a^{i\dagger} G_{i h} (E^{-1})^{h k} (E^{ T})_{k j} \tilde a^{j\dagger}
-2 a^{i\dagger} G_{i h} (E^{-1})^{h j}  z_j
-\pi \alpha' q F_{i j} \bar z ^j \tilde a^{i\dagger}
}
|0>
=
\nonumber\\
&&\hspace{1em}=
\int \prod_i \frac{d z_i ~d \bar z^i}{ \pi}
e^{- z_i \left[ \delta^i_j -2 \pi \alpha' q  F_{j h} (E^{-1})^{h i}
\right] \bar z^j }
~e^{ 2 \pi \alpha' q
\bar z^j F_{j h}(E^{-1})^{h k} G_{k i} \tilde a^{i\dagger}
}
e^{-2 a^{i\dagger} G_{i h} (E^{-1})^{h j}  z_j }
\nonumber\\
&&\hspace{4em}
e^{
-a^{i\dagger} G_{i h} (E^{-1})^{h k} (E^{ T})_{k j} \tilde a^{j\dagger}
}
|0>
=
\nonumber\\
&&\hspace{1em}=
\left[ \det ( \delta^i_j - 2\pi \alpha' q (F E^{-1})_i ^j
  \right]^{-1}
e^{-2 ~a^{i\dagger} G_{i h} (\calE^{-T})^{h k} (\hat F)_{k l}
  (E^{-1})^{l m} G_{m j}   \tilde a^{j\dagger} }
\nonumber\\
&&\hspace{4em}
e^{- ~a^{i\dagger} G_{i h} ( E^{-1})^{h k} (E^{ T})_{k j}
\tilde a^{j\dagger} }
|0>
\nonumber\\
&&\hspace{1em}=
\left[ \det (\calE^T E^{-1})_i ^j
  \right]^{-1}
e^{-
~a^{i\dagger} G_{i h} (\calE^{-T})^{h k} (\calE)_{k j} \tilde a^{j\dagger} }
|0>
\end{eqnarray*}
Then, by using the zeta function regularization $\sum_1^\infty
1=\zeta(0)=-\frac{1}{2}$, we can get the complete contribution from non
zero-mode
\begin{equation*}
\left[
Tr\left( P~e^{-i\oint q A} \right) \,
|D25(E,F=0)\rangle
\right]_{n z m}\!\!
=
\sqrt{ \det (\calE^T E^{-1})}
e^{- \sum_{n=1}^\infty
~a^{i\dagger}_n (G \calE^{-T} \calE)_{i j} \tilde a^{j\dagger_n} }
|0>
\end{equation*}
We can now examine the zero modes contribution and we find
\begin{eqnarray*}
&&
\left[
Tr\left( P~e^{-i\oint q A} \right) ~
|D25(E,F=0)\rangle
\right]_{z m}
=
\nonumber\\
&&=
\frac{\sqrt{\det E}}{( \det G)^{1/4}}
\sum_s Tr(\omega_1^{s_1}\omega_2^{s_2})~
e^{-i \pi \hat F_{12} s^1 s^2}
|n_i= \hat F_{i j} s^j, m^i=s^i\rangle
\end{eqnarray*}
The previous calculation can also be generalized to a generic torus, getting:
\begin{eqnarray*}
&&
\left[
Tr\left( P~e^{-i\oint q A} \right) ~
|D25(E,F=0)\rangle
\right]_{z m}
=
\nonumber\\
&&=
\frac{\sqrt{\det E}}{( \det G)^{1/4}}
\sum_s Tr(\omega_1^{s_1}\omega_2^{s_2}\dots\omega_{\hat d}^{s_{\hat d}})~
e^{-i \pi \hat F^<_{i j} s^i s^j}
|n_i= \hat F_{i j} s^j, m^i=s^i\rangle
\end{eqnarray*}
where  ${\hat F}^<_{i j}={\hat F}_{ij}$ if $i<j$, zero otherwise.
The factor $Tr(\omega_1^{s_1}\dots\omega_{\hat d}^{s_{\hat d}})~$ acts as a
projector on the possible values of the integers $s$.
This projector depends
explicitly on the form of the various $\omega$ but we can nevertheless
deduce the important constraints
\begin{equation*}
n_i= \hat F_{i j} s^j \in \Z
\end{equation*}
which are valid for all the values of $s$ which survive the
projection. The proof is very easy and for  $n_1$ goes as
\begin{eqnarray*}
Tr(\omega_1^{s_1}\dots\omega_{\hat d}^{s_{\hat d}})
&=&
Tr(\omega_1 \omega_1^{s_1}\dots\omega_{\hat d}^{s_{\hat d}} \omega_1^{-1})
\nonumber\\
&=&
e^{i 2\pi \hat F_{ 1 j} s^j}
Tr(\omega_1^{s_1}\dots\omega_{\hat d}^{s_{\hat d}})
\end{eqnarray*}
The final form of the boundary reads
\begin{eqnarray}
|D25(E,F)\rangle
&=&
\frac{T_{25}}{2}
~N
\frac{\sqrt{\det \calE}}{( \det G)^{1/4}}
\sum_s \frac{ Tr(\omega_1^{s_1}\omega_2^{s_2}\dots\omega_{\hat d}^{s_{\hat d}})}{N}
e^{-i \pi \hat F^<_{i j} s^i s^j}
|n_i= \hat F_{i j} s^j, m^i=s^i\rangle
\nonumber\\
&&\times
e^{-
~a^{i\dagger} G_{i h} (\calE^{-T})^{h k} (\calE)_{k j} \tilde a^{j\dagger} }
|0>
\label{naive-NA-D25}
\end{eqnarray}
with:
\begin{equation*}
 Tr(\omega_1^{s_1}\omega_2^{s_2}\dots\omega_{\hat d}^{s_{\hat d}})
=N\,
\delta^{[N]}_{s_1,0}
\dots
\delta^{[N]}_{s_{\hat d},0}
\end{equation*}

\end{document}